\documentclass[useAMS,usenatbib]{mn2e}
\usepackage{txfonts}
\usepackage{natbib}
\usepackage[all]{xy}
\usepackage[dvips]{graphicx}

\def\del#1{{}}
\def\spirou#1{{#1}}

\newcommand{\ltsima}{$\; \buildrel < \over \sim \;$}
\newcommand{\lsim}{\lower.5ex\hbox{\ltsima}}
\newcommand{\gtsima}{$\; \buildrel > \over \sim \;$}
\newcommand{\gsim}{\lower.5ex\hbox{\gtsima}}
\newcommand{\bra}{\langle}
\newcommand{\ket}{\rangle}

\newcommand{\dd}{\mathrm{d}}

\newcommand{\vecl}{\bmath{L}}
\newcommand{\vecj}{\bmath{j}}
\newcommand{\vecx}{\bmath{x}}
\newcommand{\vecr}{\bmath{r}}
\newcommand{\veck}{\bmath{k}}
\newcommand{\vecq}{\bmath{q}}

\newcommand{\vecv}{\bupsilon}
\newcommand{\vecw}{\bmath{w}}

\newcommand{\ci}{\mathrm{i}}

\newcommand{\trace}{\mathrm{tr}}
\newcommand{\determinant}{\mathrm{det}}

\newcommand{\matrixx}{\mathbfss{X}}
\newcommand{\matrixi}{\mathbfss{I}}

\title[Galactic angular momenta]
{Review: galactic angular momenta and angular momentum correlations in the cosmological large-scale structure}
\author[B.M. Sch{\"a}fer]
{Bj{\"o}rn Malte Sch\"afer\thanks{e-mail: bjoern.malte.schaefer@ias.u-psud.fr}\\
Institute of Cosmology and Gravitation, University of Portsmouth, Mercantile Building, Portsmouth PO1 2EG, United Kingdom\\
Institut d'Astrophysique Spatiale, Universit{\'e} de Paris XI, b{\^a}timent 120-121, Centre universitaire d'Orsay, 91400 Orsay CEDEX, France}

\begin{document}
\pagerange{\pageref{firstpage}--\pageref{lastpage}}
\pubyear{2008}
\maketitle
\label{firstpage}

% --- abstract --- %
\begin{abstract}
I review the theory of angular momentum acquisition of galaxies by tidal torquing, the resulting angular momentum distribution, the angular momentum correlation function and discuss the implications of angular momentum alignments on weak lensing measurements: Starting from linear models for tidal torquing I summarise perturbative approaches and the results from $n$-body simulations of cosmic structure formation. Then I continue to discuss the validity of decompositions of the tidal shear and inertia fields, the effects of angular momentum biasing, the applicability of parameterised angular momentum correlation models and the consequences of angular momentum correlations for shape alignments. I compile the result of observations of shape alignments in recent galaxy surveys as well as in $n$-body simulations. Finally, I review the contamination of weak lensing surveys by spin-induced shape alignments and methods for suppressing this contamination.
\end{abstract}

% --- keywords --- %

% --- section: introduction --- %
\section{Introduction}
This article reviews models of angular momenta of galaxies and angular momentum correlations in the cosmological large-scale structure, and the implications of angular momentum induced shape alignments of galaxies for gravitational lensing. The current paradigm states that galaxies acquire their angular momentum by a mechanism known as tidal torquing  \citep{hoyle, 1969ApJ...155..393P, 1955MNRAS.115....2S, 1970Ap......6..320D, 1984ApJ...286...38W}, where tidal fields from the ambient large-scale structure excert a torquing moment onto the protogalactic object prior to gravitational collapse. The angular momentum, which is conserved during collapse, has important implications for the formation and orientation of the galactic disk. Tidal torquing explains naturally angular momentum correlations, as neighbouring galaxies form from correlated initial conditions and experience similar tidal fields. An important application of angular momentum correlations is gravitational lensing: The angular momentum-induced ellipticity alignments are a source of systematics, which, if uncorrected, deteriorates constraints on cosmological parameters and introduces estimation biases.

After a compilation of the key formul{\ae} of cosmology, structure formation and perturbation theory in Sect.~\ref{sect_cosmology}, I review angular momentum build-up by tidal torquing, which is a second order perturbative process arising in Lagrangian perturbation theory, in Sect.~\ref{sect_torquing}. The distribution of angular momentum values and approximations used in the derivation of the angular momentum correlation function are discussed in Sects.~\ref{sect_distribution} and~\ref{sect_correlation}, respectively. Analytical derivations are contrasted with the results from $n$-body simulations of cosmic structure formation in Sect.~\ref{sect_simulation} and to observations of angular momentum induced galaxy shape correlations in Sect.~\ref{sect_observation}. Implications of such spin-induced shape correlations for weak gravitational lensing are derived in Sect.~\ref{sect_lensing}, and a summary in Sect.~\ref{sect_summary} concludes the review article.

The choice of specific values for the cosmological parameters is not as relevant for angular momentum correlations compared to other cosmological observations, because the angular momentum build-up of galaxies takes place at early times when the universe is matter dominated. Nevertheless I will attempt to keep the derivations as general as possible and use as a framework spatially flat dark energy cosmological models with Gaussian adiabatic initial perturbations in the cold dark matter density field.

% --- section: cosmology --- %
\section{cosmology and structure formation}\label{sect_cosmology}
This section provides a compilation of dark energy cosmologies (Sect.~\ref{sect_cosmology_de_cosmo}), the description of the fluctuations statistics of the density field with power spectra (Sect.~\ref{sect_cosmology_spectrum}), the structure formation equations and their linearisation (Sect.~\ref{sect_cosmology_structure}), the linear growth equation (Sect.~\ref{sect_cosmology_linear}), Lagrangian perturbation theory (Sect.~\ref{sect_cosmology_perturbation}) and halo formation by spherical collapse (Sect.~\ref{sect_cosmology_galaxy}).

% --- subsection: dark energy cosmologies --- %
\subsection{Dark energy cosmologies}\label{sect_cosmology_de_cosmo}
In a spatially flat dark energy cosmology with the matter density parameter $\Omega_m$, the Hubble function $H(a)=\dd\ln a/\dd t$ is given by
\begin{equation}
\frac{H^2(a)}{H_0^2} = \frac{\Omega_m}{a^{3}} + (1-\Omega_m)\exp\left(3\int_a^1\dd a^\prime\:\frac{1+w(a^\prime)}{a^\prime}\right),
\end{equation}
with the dark energy equation of state $w(a)$, \spirou{and omitting the radiation density $\Omega_r$, which has a negligible influence on the late-time expansion}. The value $w\equiv -1$ corresponds to the cosmological constant $\Lambda$. The relation between comoving distance $\chi$ (given in terms of the Hubble distance $d_H=c/H_0$) and scale factor $a$ is given by
\begin{equation}
\chi = c\int_a^1\dd a^\prime\:\frac{1}{{a^\prime}^2 H(a^\prime)},
\end{equation}
with the speed of light $c$. The dark energy equation of state $w(a)$ is conveniently \spirou{approximated} by its first order Taylor expansion with respect to the scale-factor $a$,
\begin{equation}
w(a) = w_0 + (1-a) w_a,
\end{equation}
introduced by \citet{2001IJMPD..10..213C} and \citet{2003MNRAS.346..573L}, \spirou{for non-interacting dark energy models with a slow time evolution $w(a)$}. The cosmic time in units of the Hubble time $t_H=1/H_0$ follows directly from the definition of the Hubble function,
\begin{equation}
t = \int_a^1\dd a^\prime\: \frac{1}{a^\prime H(a^\prime)}.
\end{equation}
For a number of purposes, in particular as the time variable in the structure formation equations, the definition of the conformal time $\eta$ in analogy to the comoving distance $\chi$ from the differential $\dd\eta=\dd t/a=\dd a/(a^2H(a))$ is useful.

% --- subsection: CDM density spectrum --- %
\subsection{CDM density spectrum}\label{sect_cosmology_spectrum}
The fluctuations of the cosmic density field $\delta(\bmath{x})$, which is defined as the fractional perturbation of the density field $\rho(\bmath{x})$ relative to the background density $\bra\rho\ket=\Omega_m\rho_\mathrm{crit}$,
\begin{equation}
\delta(\bmath{x}) = \frac{\rho(\bmath{x})-\bra\rho\ket}{\bra\rho\ket},
\end{equation}
are assumed to be Gaussian and are described by the variance between two Fourier modes $\delta(\bmath{k})=\int\dd^3x\:\delta(\bmath{x})\exp(-\ci\bmath{k}\bmath{x})$,
\begin{equation}
\bra\delta(\bmath{k})\delta(\bmath{k}^\prime)\ket=(2\pi)^3\delta_D(\bmath{k}+\bmath{k}^\prime)P(k),
\end{equation} 
which defines the power spectrum $P(k)$ in the case of homogeneous and isotropic fluctuation statistics. Inflationary models suggest the ansatz
\begin{equation}
P(k)\propto k^{n_s}T^2(k)
\end{equation}
\citep{1982PhRvL..49.1110G} for the power spectrum with the CDM transfer function $T(k)$, which is well approximated with the polynomial fit proposed by \citet{1986ApJ...304...15B},
\begin{displaymath}
T(q) = \frac{\ln(1+2.34q)}{2.34q}\left(1+3.89q+(16.1q)^2+(5.46q)^3+(6.71q)^4\right)^{-\frac{1}{4}}.
\label{eqn_cdm_transfer}
\end{displaymath}
The wave vector $k=q\Gamma$ is rescaled with the shape parameter $\Gamma$, which describes the peak shape of the CDM power spectrum $P(k)$. \spirou{\citet{1995ApJS..100..281S} gives a convenient fit to numerical data in cosmologies with $\Omega_m\neq 1$,
\begin{equation}
\Gamma=\Omega_m h\exp\left(-\Omega_b\left(1+\frac{\sqrt{2h}}{\Omega_m}\right)\right),
\end{equation}
where $\Gamma$ is measured in units of $(\mathrm{Mpc}/h)^{-1}$, such that $q$ is dimensionless}. The normalisation is given by the variance $\sigma_8$ of the density field on the scale $R=8~\mathrm{Mpc}/h$ at the current cosmic epoch,
\begin{equation}
\sigma^2_R = \frac{1}{2\pi^2}\int\dd k\:k^2 P(k) W^2(kR),
\end{equation}
with a Fourier transformed spherical top hat filter function, $W(x)=3j_1(x)/x$. $j_\ell(x)$ is the spherical Bessel function of the first kind of order $\ell$  \citep{1972hmf..book.....A, 2005mmp..book.....A}.

% --- subsection: structure formation with CDM --- %
\subsection{Structure formation with CDM}\label{sect_cosmology_structure}
The formation of cosmic structures is based on gravitational amplification of seed perturbations in the cosmic density field. Being a hydrodynamical self-gravitating phenomenon, structure formation is described in the comoving frame by the system of differential equations composed of $(i)$ the continuity equation
\begin{equation}
\frac{\partial}{\partial\eta}\delta + \mathrm{div}\left[(1+\delta)\vecv\right] = 0,
\end{equation}
which relates the time-evolution of the density field to the divergence of the matter fluxes $\vecj=(1+\delta)\vecv$, $(ii)$ the Euler-equation
\begin{equation}
\frac{\partial}{\partial\eta}\vecv + aH\vecv + \vecv\nabla\vecv = -\nabla\Phi,
\end{equation}
which describes the evolution of the peculiar velocity field $\vecv$ from the \spirou{gradient $\nabla\Phi$ of the peculiar gravitational potential $\Phi$,} acting on a fluid element, and finally $(iii)$ the Poisson-equation
\begin{equation}
\Delta\Phi = \frac{3}{2}\Omega_m(\eta)\:(aH)^2\delta,
\end{equation} 
which gives the gravitational potential $\Phi$ induced by the matter distribution $\delta$ (Newton's constant has been replaced with the definition of the critical density, and absorbed into the ambient density into $\Omega_m(\eta)$ with the conformal time $\eta$ as the time variable). The expanding background is described by the Hubble function $H(a)$ and the usage of Newtonian dynamics and Newtonian gravity is well justified on sub-horizon scales. Due to the fact that cold dark matter is collisionless, viscous and pressure forces are negligible and the formation of objects takes place by phase-space mixing, dynamical friction \citep{1943ApJ....97..255C} and violent relaxation \citep{1967MNRAS.136..101L}.

Linearisation of the structure formation equations by substituting a perturbative expansion of the density- and velocity field yields the linearised continuity equation,
\begin{equation}
\frac{\partial}{\partial\eta}\delta + \mathrm{div}\vecv = 0,
\end{equation}
and the linearised Euler-equation,
\begin{equation}
\frac{\partial}{\partial\eta}\vecv + aH\vecv = -\nabla\Phi,
\end{equation}
which are valid if the overdensity $\delta$ is small, $\delta\ll 1$. 

It is instructive to decompose the velocity field into the divergence $\theta=\mathrm{div}\vecv$ and the vorticity $\bomega=\mathrm{rot}\vecv$, for which one obtains the evolution equations from the linearised Euler-equation
\begin{equation}
\frac{\partial}{\partial\eta}\theta + aH\theta + \frac{3}{2} \Omega_m(\eta)\: (aH)^2\delta = 0
\quad\mathrm{and}\quad
\frac{\partial}{\partial\eta}\bomega + aH\bomega = 0.
\end{equation}
Solving the latter equation shows that the vorticity $\bomega$ decreases $\propto 1/a$ such that any initial vortical excitation is wiped out rapidly, which disfavours models in which galactic angular momenta arise from vortical initial perturbations in the velocity field \citep{1970Ap......6..203Z, 1976RvMP...48..107J, 1977ARA&A..15..235G, 1983FCPh....9....1E}. Vorticity can then only be generated by a nonlinear amplification term of the form $\mathrm{rot}(\bupsilon\times\bomega)$ in the full Euler-equation if there are non-zero stresses. The flows which develop in a collisionless fluid are necessarily laminar potential flows. Only collective processes between the dark matter particles can emulate viscosity and pressure, which happens on fully nonlinear scales, and generate vorticity by the nonlinear term $\mathrm{rot}(\bupsilon\times\bomega)$ \citep{1999A&A...343..663P, 2002PhR...367....1B}.

% --- subsection: linear growth --- %
\subsection{Linear structure growth}\label{sect_cosmology_linear}
The linearised continuity equation can be combined with the time derivative of the evolution equation for the velocity divergence to form the growth equation, whose solution describes the homogeneous growth $\delta(\bmath{x},a)=D_+(a)\delta(\bmath{x},a=1)$ of the density field in the linear regime of structure formation, $\left|\delta\right|\ll 1$ \citep{1980PhyS...21..720P, 1997PhRvD..56.4439T, 2003MNRAS.346..573L},
\begin{equation}
\frac{\dd^2}{\dd a^2}D_+(a) + \frac{1}{a}\left(3+\frac{\dd\ln H}{\dd\ln a}\right)\frac{\dd}{\dd a}D_+(a) = 
\frac{3}{2a^2}\Omega_m(a) D_+(a).
\label{eqn_growth}
\end{equation}
The growing solution $D_+(a)$ of the growth equation's two solutions is referred to as the growth function and assumes the simple solution $D_+(a)=a$ in the SCDM cosmology, where $\Omega_m\equiv 1$ and $H(a)/H_0 = a^{-3/2}$. This scaling motivates the usage of the scale factor $a$ as the preferred time variable, and to transform the differentials with $\dd/\dd\eta = a^2H(a)\dd/\dd a$. Hence, the initial conditions for solving the growth equation are given by $D_+(0) = 0$ and $\dd D_+(0)/\dd a = 1$, due to matter domination in early times. The term $3+\dd\ln H/\dd\ln a$ is equal to $2-q$, with the deceleration parameter $q$, and acts as a fiction term which suppresses structure formation on scales whose dynamical time scale is larger than the time scale of the Hubble expansion. 

Homogeneous structure formation is equivalent to the notion of independently growing Fourier modes,
\begin{equation}
\delta(\bmath{x},a) = D_+(a)\delta(\bmath{x},a=1) \longrightarrow\delta(\bmath{k},a) = D_+(a)\delta(\bmath{k},a=1),
\label{eqn_homo_growth}
\end{equation}
which conserves the Gaussianity of the initial conditions. \spirou{The Gaussianity of the initial density perturbations is a consequence of inflation, where a large number of uncorrelated quantum flucutations are superimposed, yielding a Gaussian amplitude distribution due to the central limit theorem. As long as the growth of structure is linear and thus homogeneous as described by eqn.~(\ref{eqn_homo_growth}), all modes growth independently conserving the Gaussianity of the initial density field.}

% --- subsection --- %
\subsection{Lagrangian perturbation theory and nonlinear growth}\label{sect_cosmology_perturbation}
The appropriate tool for describing the angular momentum build-up of haloes in the large-scale structure, which is a perturbative process, is Lagrangian perturbation theory \citep[for an detailed review, see][]{1995PhR...262....1S, 2002PhR...367....1B}. The central objects in Lagangian perturbation theory are the particle trajectories linking the initial positions $\vecq$ to their positions $\vecx$ at time $\eta$,
\begin{equation}
\vecx(\vecq,\eta) = \vecq - \nabla\Psi(\vecq,\eta),
\end{equation}
with the displacement potential $\Psi$ which describes the potential flows developing in the large-scale structure. The equation of motion for a particle is given by
\begin{equation}
\frac{\dd^2}{\dd \eta^2}\vecx + aH\frac{\dd}{\dd\eta}\vecx = -\nabla\Phi.
\end{equation}
The divergence of this equation yields the connection to the Eulerian approach by relating the physical gravitational potential $\Phi$ to the displacement potential $\Psi$ and allows to write by using the Poisson equation
\begin{equation}
J(\vecq,\eta)\:\mathrm{div}\left[\frac{\dd^2}{\dd\eta^2} + aH\frac{\dd}{\dd\eta}\right]\nabla\Psi = \frac{3}{2}\Omega_m(\eta)(aH)^2(J(\vecq,\eta)-1),
\end{equation}
with the mass conservation condition $(1+\delta)\dd^3x=\dd^3q$. Therefore, $1+\delta(\vecx,\eta) = 1/J(\vecq,\eta)$ with the Jacobian 
\begin{equation}
J(\vecq,\eta)\equiv 
\frac{\dd\vecx}{\dd\vecq} = 
\left(\mathrm{det}\left[\delta_{ij} + \partial_i\partial_j\Psi\right]\right)^{-1},
\end{equation} 
meaning that the tidal forces $\partial_i\partial_j\Psi$ cause a compression of the cosmic density field. The limit of applicability of Lagrangian perturbation theory is reached if $J=0$, because the density field formally diverges. This happens if two trajectories cross and if the mapping $\vecq\rightarrow\vecx$ becomes ambiguous. The linear solution to $\Psi$ can be derived as $\Delta\Psi(\vecq,\eta) = D_+(\eta)\delta(\vecq)$, using the solution $D_+(\eta)$ to the homogeneous growth equation \citep{1995MNRAS.276..115C}. With this potential and the perturbative mapping $\vecx(\vecq,\eta) = \vecq - \nabla\Psi(\vecq,\eta)$ is refered to as the Zel'dovich-approximation \citep{1970A&A.....5...84Z, 1992MNRAS.254..729B}. The analytical power in the Zel'dovich approximation lies in the extrapolation of nonlinear dynamics from the initial conditions, without having to follow the evolution of nonlinearities computationally in the Eulerian approach \citep{1992ApJ...394L...5B, 1993MNRAS.264..375B, 1995A&A...296..575B, 1995PhR...262....1S, 1995MNRAS.276...39C, 2002PhR...367....1B}.

% --- subsection: galaxy formation --- %
\subsection{Galaxy formation and biasing}\label{sect_cosmology_galaxy}
Galaxies form at local peaks in the large scale structure by ellipsoidal collapse \citep{2001MNRAS.323....1S} of the dark matter perturbation. In the collapse, the halo decouples from the local Hubble expansion which is overcome by the halos own gravitational field. The baryonic component inside a dark matter halo undergoes radiative cooling and forms a rotationally supported galactic disk. 

The fractional perturbation $\Delta n/n$ in the mean number density $n$ of dark matter haloes hosting a galaxy is related to the overdensity field $\delta=\Delta\rho/\bra\rho\ket$ by the bias parameter $b$,
\begin{equation}
\frac{\Delta n}{n} = b\frac{\Delta\rho}{\bra\rho\ket},
\end{equation}
which is in general scale dependent \citep{lumsden98} and slowly time-evolving \citep{fry96, tegmark98}.

% --- section: tidal torquing ---%
\section{tidal torquing}\label{sect_torquing}
This section treats tidal torquing as a mechanism for angular momentum build-up of dark matter haloes in detail. After a short qualitative description (Sect.~\ref{sect_torquing_mechanism}), I introduce tidal torquing in Lagrangian perturbation theory using the Zel'dovich approximation as a dynamical model (Sect.~\ref{sect_torquing_zeldovich}) and discuss the time evolution of the angular momentum for constant torques in various cosmologies (Sect.~\ref{sect_torquing_time}). A crucial technical point is the misalignment between the eigensystems of the inertia tensor and the tidal shear, which I elaborate on in Sect.~\ref{sect_torquing_misalignment}, followed by a derivation of the scaling of the angular momentum with halo mass (Sect.~\ref{sect_torquing_mass_scaling}).

% --- subsection: outline --- %
\subsection{Mechanism of tidal torquing}\label{sect_torquing_mechanism}
The current paradigm for generating the angular momentum of galaxies is tidal torquing, where the tidal gravitational fields exerts a torquing moment on the protogalactic object prior to collapse. It should be emphasised that in this picture the angular momentum does {\em not} originate from the vorticity $\bomega$ which is amplified by the nonlinear term $\mathrm{rot}(\bupsilon\times\bomega)$ in the Euler-equation as explained in Sect.~\ref{sect_cosmology_structure}, but is rather generated from (vorticity-free) shear flows, in which protogalactic objects are embedded. During the shearing, each protohalo is deformed and acquires a rotational motion component. Eventually it decouples from the shear flow by collapsing under its own gravity, which reduces the length of the lever arms and makes torquing inefficient. For the perturbative description of the deformation of the object by the action of shear flows, Lagrangian perturbation theory is suitable. The difficulty in this approach lies in the fact that the shear flows themselves influence the boundary surface of the embedded object and hence its total mass and inertia. Tidal torquing effectively solves the problem of generating vorticity from a laminar flow by introducing a non-simply connected density and velocity fields, because during spherical collapse the protohalo volume is separated from the ambient fields.

This mechanism was first quantitatively investigated by \citet{1970Ap......6..320D}, \citet{1984ApJ...286...38W} and \citet{1985A&A...151..105W}, building on the original idea by \citet{hoyle} and \citet{1955MNRAS.115....2S}. Assuming a non-spherical shape of the protogalactic region, the angular momentum grows at first order and linearly in time in Einstein-de~Sitter universes, whereas in spherical regions, the acquisition of angular momentum is only a second order effect due to convective matter streams on the boundary surface, as shown by \citet{1969ApJ...155..393P}.

% --- subsection: L-acquisition by tidal torquing --- %
\subsection{Acquisition of angular momentum by tidal torquing}\label{sect_torquing_zeldovich}
Quite generally, the angular momentum $\vecl$ of a rotating mass distribution $\rho(\vecr,t)$ contained in the physical volume $V$ is given by:
\begin{equation}
\vecl(t) = \int_V\dd^3r\:\left(\vecr-\bar{\vecr}\right)\times\vecv(\vecr,t)\rho(\vecr,t),
\end{equation}
where $\vecv(\vecr,t)$ is the (rotational) velocity of the fluid element with density $\rho(\vecr,t) = \bra\rho\ket(1+\delta(\vecr,t))$ at position $\vecr$ around the centre of gravity $\bar{\vecr}$. In perturbation theory, $\delta\ll 1$, and the density field can be approximated by assuming a constant density $\bra\rho\ket = \Omega_m\rho_\mathrm{crit}$ inside the protogalactic region. \citet{1984ApJ...286...38W}, \citet{1996MNRAS.282..436C}, \citet{1997ASPC..117..431T} and \citet{2001ApJ...559..552C} have described the growth of perturbations on an expanding background in Lagrangian perturbation theory: The trajectory of dark matter particles in comoving coordinates to first order is given by the Zel'dovich approximation as the dynamical model \citep{1970A&A.....5...84Z}:
\begin{equation}
\vecx(\vecq,t) = \vecq - D_+(t)\nabla\Psi(\vecq)
\rightarrow
\dot{\vecx} = -\dot{D}_+\nabla\Psi,
\end{equation}
which relates the initial particle positions $\vecq$ to the positions $\vecx$ at time $t$. The particle velocity $\dot{\vecx}$ follows from the Zel'dovich-relation by differentiation by the time-variable. The growth function $D_+(t)$ describes the homogeneous time evolution of the displacement field $\Psi$ and contains the influence of the particular dark energy model. In the Lagrangian frame, the expression for the angular momentum becomes
\begin{equation}
\vecl 
= \rho_0 a^5 \int_{V_L}\dd^3q\:\left(\vecx-\bar{\vecx}\right)\times \dot{\vecx}
\simeq \rho_0 a^5 \int_{V_L}\dd^3q \left(\vecq - \bar{\vecq}\right)\times \dot{\vecx},
\end{equation}
where the integration volume is defined in comoving coordinates as well. 
Assuming that the gradient $\nabla\Psi(\vecq)$ of the displacement field $\Psi(\vecq)$ does not vary much across the Lagrangian volume $V_L$, a second-order Taylor expansion in the vicinity of the centre of gravity $\bar{\vecq}$ is applicable:
\begin{equation}
\partial_\alpha\Psi(\vecq) \simeq \partial_\alpha\Psi(\bar{\vecq}) + \sum_\beta (\vecq - \bar{\vecq})_\beta\Psi_{\alpha\beta}, 
\end{equation}
The expansion coefficient is the tidal shear $\Psi_{\sigma\gamma}$ at the point $\bar{\vecq}$:
\begin{equation}
\Psi_{\sigma\gamma}(\bar{\vecq}) = \partial_\sigma\partial_\gamma\Psi(\bar{\vecq}),
\end{equation}
because the Zel'dovich displacement field $\Psi$ is related to gravitational potential $\Phi$ and can be computed as the solution to Poisson's equation $\Delta\Psi=\delta$ from the cosmological density field $\delta$. The gradient $\partial_\alpha\Psi(\bar{\vecq})$ of the Zel'dovich potential displaces the protogalactic object, which is neglected in the further derivation, as I only trace differential advection velocities responsible for inducing rotation. Identifying the tensor of second moments of the mass distribution of the protogalactic object as the inertia $I_{\beta\sigma}$,
\begin{equation}
I_{\beta\sigma} = \rho_0 a^3 \int_{V_L}\dd^3q\: (\vecq - \bar{\vecq})_\beta (\vecq -\bar{\vecq})_\sigma
\end{equation}
one obtains the final expression of the angular momentum $L_\alpha$:
\begin{equation}
L_\alpha = a^2 \dot{D}_+ \epsilon_{\alpha\beta\gamma} \sum_\sigma I_{\beta\sigma}\Psi_{\sigma\gamma}.
\label{eqn_l_definition}
\end{equation}
\spirou{Here, eqn.~(\ref{eqn_l_definition}) implicitly assumes a peak constraint because the inertia of the protogalactic volume is needed, which can only be defined in a sensible way for a density peak from which a halo forms by collapse.}

\begin{figure}
\resizebox{\hsize}{!}{\includegraphics{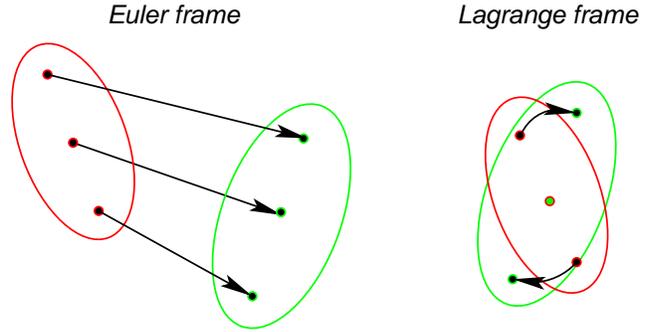}}
\caption{Principle of tidal torquing: A linear variation of the displacements across the protogalactic object in the space-fixed Euler-frame translates to a rotational motion in the comoving Lagrange-frame \spirou{after collapse}.}
\label{fig_tidal_torquing}
\end{figure}

Fig.~\ref{fig_tidal_torquing} illustrates the torquing action on a protogalactic region in the Euler- and Lagrange frames: If the Zel'dovich displacements $\nabla\Psi$ vary linearly across the protogalactic cloud, they introduce a rotational motion in the comoving Lagrangian frame. Being proportional to the second derivative of the potential $\Psi$, the rotation is determined by the tidal forces $\Psi_{\sigma\gamma}=\partial_\sigma\partial_\gamma\Psi$. Tidal torquing is effective until the moment of turn-around in the spherical collapse picture, because the collapse dramatically reduces the lever arms. After the collapse, the halo conserves the angular momentum it has accumulated until turn-around.

\citet{2002MNRAS.332..325P} give an intuitive argument on the orientation of the angular momentum relative to the principal axis system of the tidal shear, which is a very useful relation: In the eigenframe of the tidal shear, the components of the angular momentum are simply given by $L_\alpha\propto I_{\beta\gamma}(\Psi_\beta-\Psi_\gamma)$, where the indices are cyclic permutations of $(1,2,3)$. $\Psi_\alpha$ with a single index are the three eigenvalues of $\bmath{\Psi}$. Averaging over all orthogonal transformations relating the eigensystems of $\matrixi$ and $\bmath{\Psi}$ yields the result that the largest component of $\vecl$ would be the one for which $\left|\Psi_\beta-\Psi_\gamma\right|$ is largest. Because of the ordering $\Psi_1\leq\Psi_2\leq\Psi_3$, this is necessarily $L_2$, $L_2\propto\left|\Psi_3-\Psi_1\right|$, i.e. the angular momentum tends to align itself parallel to $\Psi_2$, i.e. the axes corresponding to the intermediate eigenvalue of the tidal shear. \spirou{This argument is not generalisable to the case with correlations between shear and inertia: In the ideal case of a Gaussian random field, one can transform into the eigensystem of the shear $\bmath{\Psi}$, such that only the off-diagonal components of $\matrixi$ are used in the computation of $\vecl$, which are statistically uncorrelated, and this argument fails if there are contributions from the diagonal components of $\matrixi$, which can disturb the ordering of the angular momentum components.}

% --- subsection: time evolution --- %
\subsection{Time evolution of the angular momentum}\label{sect_torquing_time}
\begin{figure}
\resizebox{\hsize}{!}{\includegraphics{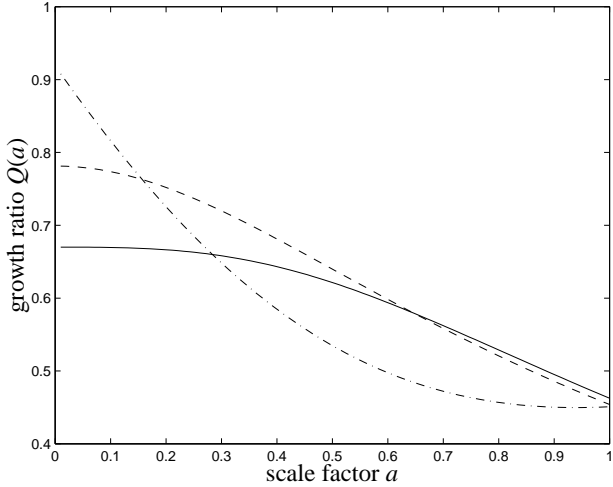}}
\caption{The influence of the dark energy model on the time evolution of angular momenta: $Q(a)$ as a function of scale factor $a$ for $\Lambda$CDM ($w_0=-1$ and $w_a=0$, solid line), for quintessence ($w_0=-2/3$ and $w_a=0$, dashed line) and a dark energy model with variable equation of state ($w_0=-2/3$ and $w_a=1/3$, dash-dotted line). The matter density parameter was chosen to be $\Omega_m=0.25$ in all cases.}
\label{fig_time_evolution}
\end{figure}

For investigating the influence of dark energy on the spin-up via the $\dot{D}_+$-factor, it is convenient to rewrite the time dependence of $D_+$ in terms of the scale factor $a$ by $\dd D_+/\dd t = a H(a)\dd D_+/\dd a$, yielding:
\begin{equation}
L_\alpha = a^3 H(a)\frac{\dd D_+}{\dd a}\:
\epsilon_{\alpha\beta\gamma} \sum_\sigma I_{\beta\sigma}\Psi_{\sigma\gamma}.
\end{equation}
Fig.~\ref{fig_time_evolution} compares the time evolution of the angular momentum in dark energy cosmologies with SCDM. I define the ratio
\begin{equation}
Q(a)\equiv \frac{q_\mathrm{DE}(a)}{q_\mathrm{SCDM}(a)}
\end{equation}
with 
\begin{equation}
q(a)=a^3 H(a) \frac{\dd D_+}{\dd a},
\end{equation}
where $D_+(a)$ is normalised to unity today. In SCDM these formul{\ae} simplify to $H(a)/H_0=a^{-3/2}$, $D_+(a)=a$ and consequently $q_\mathrm{SCDM}=H_0a^{3/2}$. Fig.~\ref{fig_time_evolution} compares $Q(a)$ for dark energy models with a constant and a time-evolving equation of state to a model with a cosmological constant $\Lambda$ and suggests that the spin-up of haloes in dark energy models half as fast compared to SCDM, and the choice of the equation of state affects the time evolution significantly \citep{my_paper}. 

% --- subsection: misalignment --- %
\subsection{Misalignment of the shear and inertia eigensystems}\label{sect_torquing_misalignment}
In order to illustrate the consequence of the contraction of the product $\matrixx=\mathbfss{I}\bmath{\Psi}$ with the Levi-Civit\`{a}-symbol $\epsilon_{\alpha\beta\gamma}$ in the definition of the angular momentum, one can carry out a decomposition of this tensor, $\matrixx = \matrixx^+ + \matrixx^-$ \citep{my_paper}, into an antisymmetric contribution $\matrixx^-$, defined via the commutator $\left[\mathbfss{I},\bmath{\Psi}\right]$,
\begin{equation}
\matrixx^- \equiv \frac{1}{2}\left[\mathbfss{I},\bmath{\Psi}\right], \quad
X_{\beta\gamma}^- = \frac{1}{2}\sum_\sigma\left(I_{\beta\sigma}\Psi_{\sigma\gamma} - \Psi_{\beta\sigma}I_{\sigma\gamma}\right),
\end{equation}
with the symmetry $(\matrixx^-)^t = \frac{1}{2}(\mathbfss{I}\bmath{\Psi}-\bmath{\Psi}\mathbfss{I})^t = \frac{1}{2}(\bmath{\Psi}\mathbfss{I} - \mathbfss{I}\bmath{\Psi}) = - \matrixx^-$ under matrix transposition ($\mathbfss{I}$ and $\bmath{\Psi}$ are symmetric matrices) and into the corresponding symmetric matrix $\matrixx^+$ by using the anticommutator $\left\{\mathbfss{I},\bmath{\Psi}\right\}$ between inertia $\mathbfss{I}$ and tidal shear $\bmath{\Psi}$:
\begin{equation}
\matrixx^+ 
\equiv \frac{1}{2}\left\{\mathbfss{I},\bmath{\Psi}\right\},\quad
X_{\beta\gamma}^+ = \frac{1}{2}\sum_\sigma\left(I_{\beta\sigma}\Psi_{\sigma\gamma} + \Psi_{\beta\sigma}I_{\sigma\gamma}\right),
\end{equation}
with $(\matrixx^+)^t = +\matrixx^+$. In the derivation of the angular momentum $\vecl$,
\begin{equation}
L_\alpha 
= a^2 \dot{D}_+ \epsilon_{\alpha\beta\gamma}\sum_\sigma I_{\beta\sigma}\Psi_{\sigma\gamma}
= a^2 \dot{D}_+ \epsilon_{\alpha\beta\gamma} X_{\beta\gamma},
\end{equation}
the permutation symbol $\epsilon_{\alpha\beta\gamma}$ picks out the antisymmetric contribution $\matrixx^-$, by virtue of $\epsilon_{\alpha\beta\gamma}(X_{\beta\gamma}^+ + X_{\beta\gamma}^-) = X_{\beta\gamma}^-$, because the contraction of the symmetric tensor $\matrixx^+$ with the antisymmetric permutation symbol vanishes, $\epsilon_{\alpha\beta\gamma} X_{\beta\gamma}^+=0$. 

Hence, the protogalactic objects will only acquire angular momentum if the commutator $\matrixx^-$ between the inertia and the tidal shear is non-zero, which means that the inertia and shear tensors are not supposed to be simultaneously diagonisable, i.e. they are not allowed to have a common eigensystem and have to be skewed relative to each other. An extreme example of the dependence of the angular momentum on the commutator $\matrixx^-$ is a spherical perturbation, which will never start rotating irrespective of the strength of the tidal fields because one can always choose the eigensystem of the inertia to coincide with that of the tidal shear, resulting in a vanishing commutator. In order to capture this mechanism, \citet{2000ApJ...532L...5L} and \citet{2001ApJ...559..552C} have used an effective, parameterised description of the average misalignment of the shear and inertia eigensystems, gauged with numerical $n$-body data. 

The symmetric contribution $\matrixx^+$, which measures the degree of alignment of the inertia and shear eigensystems, causes an anisotropic deformation of the protogalactic region during the course of linear structure formation prior to gravitational collapse. Consequently, the determination of ellipticity distributions is likely to be affected even in the stage of linear structure formation, and predictions of triaxiality based on peak shapes in Gaussian random fields \citep{1986ApJ...304...15B} could be refined using an adaptation of the formalism outlined above. Additionally, the tidal forces described by $\matrixx^+$ have an influence on the total mass of the object which will form by ellipsoidal collapse on a particular peak \citep[see e.g. ][]{2005ApJ...632..706L}. In a very detailed study, \citet{1992ApJ...401..441D} investigated the impact on tidal shearing on halo formation in an $n$-body simulation. He found significant deviations from the shapes of peaks in the initial density field compared to the halo triaxiality distribution, by the action of $\matrixx^+$, and influence of the tidal forces on the boundary defining the dark matter object which forms at a peak in the density field.

% --- subsection: scaling of angular momentum with halo mass --- %
\subsection{Scaling of angular momentum with halo mass}\label{sect_torquing_mass_scaling}
In the CDM paradigm, dark matter haloes form by spherical collapse, which decouples the dark matter object from the tidal forces. Hence, the angular momentum built up until the moment of turn around (at which the radius stalls) can be expected to be a good estimate of the total angular momentum. \spirou{As shown by \citet{1969ApJ...155..393P} for haloes in an SCDM cosmology, more massive objects have larger angular momenta, the scaling being given by $L\propto M^{5/3}$. In a low-density cosmology like $\Lambda$CDM}, the overdensity at turn around is only slightly larger than unity, $\delta=1.07$, such that the continuity equation can be written as $\delta=-D_+\Delta\Psi$, which is equivalent to setting $D_+\simeq 1/\Delta\Psi$ at turn around. \citet{1996MNRAS.282..455C} and \citet{2000MNRAS.311..762S} continue by denoting the total mass of the protohalo by $M$ and its radius at turn around by $R$, and deduct that at turn around
\begin{equation}
L\simeq a^2\dot{D}_+\Delta\Psi\:MR^2 
\simeq\frac{\dot{D}_+}{D_+}\rho^{-2/3}M^{5/3} 
\propto \Omega_m^{-0.07}\left(\frac{\dot{a}}{a}\right)^{-1/3}M^{5/3},
\end{equation}
which gives the scaling $L\propto M^{5/3}$ for the relation between angular momentum and halo mass. The term $MR^2$ is an estimate of the inertia of the protohalo. In summary, the origin of the mass-angular momentum relation is the fact that the angular momentum can grow until the time of turn-around, which is determined by the halo mass.

% --- section: angular momentum distribution --- %
\section{angular momentum distribution}\label{sect_distribution}
This section deals with the statistics of the angular momentum distribution. I provide a formal derivation of the moments of the angular momentum distribution from a Gaussian random process, restricted to galaxy formation sites (Sect.~\ref{sect_distribution_moments}), followed by a derivation of the angular momentum variance, which uses the distribution of protohalo shapes in a Gaussian random field as an input for the inertia distribution (Sect.~\ref{sect_distribution_peak}). The restriction of galaxy formation to local maxima in the density field results in biased, non-Gaussian angular momentum distributions, which are discussed in detail (Sect.~\ref{sect_distribution_bias}). Finally, the angular momentum statistics is extended to nonlinear structure formation (Sect.~\ref{sect_distribution_nonlinear}), followed by a compilation of properties of the spin-parameter $\lambda$ and its influence on disk formation (Sect.~\ref{sect_distribution_spin}).

% --- subsection: Moments of the angular momentum distribution --- %
\subsection{Moments of the angular momentum distribution}\label{sect_distribution_moments}
The probability density $p(L)\dd L$ of the galactic angular momenta $\vecl$ follows from a joint Gaussian random process, which returns the variables which are necessary to describe the angular momentum acquisition, restricted to local peaks in the large-scale structure because galaxy formation is thought to take place at local peaks only.  These variables are density field $\delta(\vecx)$, the density gradient $\delta_\alpha(\vecx)$, the second derivatives $\delta_{\alpha\beta}(\vecx)$ and the tidal field $\Psi_{\alpha\beta}(\vecx)$:
\begin{eqnarray}
\delta(\vecx) & = & 
\int\frac{\dd^3k}{(2\pi)^3}\delta(\veck)\exp(\ci\veck\vecx),\\
\delta_\alpha(\vecx) & = & 
\frac{\partial\delta(\vecx)}{\partial x_\alpha} = 
\ci\int\frac{\dd^3k}{(2\pi)^3}k_\alpha\delta(\veck)\exp(\ci\veck\vecx),\\
\delta_{\alpha\beta}(\vecx) & = & 
\frac{\partial^2\delta(\vecx)}{\partial x_\alpha\partial x_\beta} = 
-\int\frac{\dd^3k}{(2\pi)^3} k_\alpha k_\beta\delta(\veck)\exp(\ci\veck\vecx)\label{eqn_link_inertia_delta},
\end{eqnarray}
The tidal shear follows from the solution of the Poisson equation $\Delta\Psi(\vecx) = \delta(\vecx)$ linking the Zel'dovich displacement potential $\Psi(\vecx)$ to the density field $\delta(\vecx)$:
\begin{equation}
\Psi_{\alpha\beta}(\vecx) =
\frac{\partial^2\Psi(\vecx)}{\partial x_\alpha\partial x_\beta} = 
\int\frac{\dd^3k}{(2\pi)^3} \frac{k_\alpha k_\beta}{k^2} \delta(\veck)\exp(\ci\veck\vecx).
\label{eqn_link_shear_delta}
\end{equation}
An important consequence of eqns.~(\ref{eqn_link_inertia_delta}) and~(\ref{eqn_link_shear_delta}) will be the fact that the angular momentum correlation is determined by two mechanisms with differing correlation length: a short range correlation of the peak shapes and hence the inertia, and a long range correlation mediated by the tidal shear. There is an interesting analogy to solid state physics: Electron spins in ferrimagnetic materials are subjected to interactions of two different correlations lengths, too. \spirou{The occurrence of two correlation lengths is crucial to the angular momentum generation. The longer correlation length of the tidal shear ensures that the tidal field has contributions from the ambient density field, and is not sourced by the perturbation alone. If the latter were the case, shear and inertia would be degenerate and the protohalo could not acquire angular momentum.}

A peculiarity worthwhile mentioning is the fact that the density field $\delta(\vecx)$ is degenerate with the trace $\trace\Psi_{\alpha\beta}(\vecx)$ of the tidal shear because of Poisson's equation $\Delta\Psi = \trace\Psi_{\alpha\beta}=\delta$ \citep{1988MNRAS.232..339H}. For that reason, the density field will appear in the above outlined random process as a derived quantity, whereas the entries of the tidal shear matrix will be drawn from the Gaussian distribution, with the peak restriction in place. As a consequence, the multivariate Gaussian distribution has only 15 degrees of freedom, instead of 16. A second reason why this degeneracy is sensible is the fact that the trace of the tidal shear, being a coordinate system independent quantity, transforms as a scalar just like the density field itself.

The peak restriction is modeled by the criterion that the Gaussian process is required to exceed a threshold $\nu$ in density which corresponds to a minimal peak height for galaxy formation, $\delta(\vecx)>\nu$, to be of vanishing density gradient, $\delta_\alpha(\vecx)=0$, and of negative curvature, $\delta_{\alpha\beta}(\vecx)<0$. In addition, the curvature tensor $\delta_{\alpha\beta}(\vecx)<0$ is used to derive the inertia of the protogalactic region.

The number density of maxima in the density field can be modeled by a point-process \citep{1986ApJ...304...15B, 1995MNRAS.272..447R, 1999MNRAS.310.1062H}:
\begin{equation}
n_\mathrm{peak} = \sum_i\delta_D^3(\vecx-\vecx_i).
\end{equation}
Close to the maximum $i$ at $\vecx_i$ a Taylor expansion of the density gradient $\delta_\alpha(\vecx)$ is applicable:
\begin{equation}
\delta_\alpha(\vecx) = \sum_\beta\delta_{\alpha\beta}(\vecx_i)(\vecx-\vecx_i)_\beta.
\end{equation}
With this expansion on obtains for the peak density:
\begin{equation}
n_\mathrm{peak} 
= \sum_i\delta_D^3\left(\delta_{\alpha\beta}^{-1}(\vecx_i)\delta_\alpha\right) 
= \left|\determinant\:\delta_{\alpha\beta}\right|\:\delta_D^3(\delta_\alpha).
\end{equation}
The relation takes a simpler shape when considering the eigensystem of the mass tensor $-\delta_{\alpha\beta}$: Being a symmetric tensor, it has the three real eigenvalues $\lambda_i$, $i=1\ldots3$, which allows to replace the determinant by the invariant quantity $\left|\lambda_1\lambda_2\lambda_3\right|$.

The constraints can be combined in a mask $\mathcal{C}(\vecv)$, which is defined as a function on the vector $\vecv$ containing the derivates of the Gaussian random field under consideration: Peaks in the density field are defined as points with amplitudes in units of the variance $\sigma_0^2=\bra\delta^2\ket$ exceeding a certain threshold $\nu$ and exhibiting a vanishing gradient $\delta_\alpha$ as well as negative curvature $\delta_{\alpha\beta}$:
\begin{equation}
\mathcal{C}(\vecv) = 
\delta^3_D\left[\delta_\alpha(\vecx)\right]\:\left|\lambda_1\lambda_2\lambda_3\right|\:\Theta(\lambda_i)\:
\Theta\left[\delta(\vecx)-\sigma_0\nu\right].
\end{equation}
$\Theta(x)$ denotes the Heaviside step-function. By arranging all random derivates in a vector $\vecv$, the peak density $n_\mathrm{peak}$, i.e. the expecation value for the number density of peaks in the fluctuating density field $\delta$ which exceed a threshold $\nu\sigma_0$ can then be derived from the multivariate Gaussian random process $p(\vecv)\dd\vecv$,
\begin{equation}
n_\mathrm{peak} =\int\dd\vecv\: p(\vecv)\:\mathcal{C}(\vecv),
\end{equation}
which corresponds to the integral of the differential peak density $n_\mathrm{peak}(\nu)\dd\nu$ defined by \citet{1986ApJ...304...15B}.

In analogy, the moments of the angular momentum distribution with a restriction to peak regions in the density field can be obtained with \citep{my_paper}:
\begin{equation}
\bra L^n\ket = \frac{1}{n_\mathrm{peak}(>\nu)}\int\dd\vecv\: p(\vecv)\: \mathcal{C}(\vecv) L^n(\vecv).
\label{eqn_l_1pt_variance}
\end{equation}
Under the assumption of isotropy, it is sufficient to consider the distribution $p(L)\dd L$ of the angular momentum along a single dimension. In both formul{\ae}, the normalisation factor $1/n_\mathrm{peak}$ accounts for the discreteness of the measured quantity. In principle, all peak restricted moments $\bra L^n\ket$ can be derived and the distribution $p(L)\dd L$ reconstructed using the characteristic function $\varphi(t)$ \citep{2005mmp..book.....A},
\begin{equation}
\varphi(t)=\int\dd L\: p(L)\exp(-\ci Lt) = \sum_n\bra L^n\ket\frac{(-\ci t)^n}{n!},
\end{equation}
with $\bra L^n\ket=\int\dd L\:L^n p(L)$, and successive inverse Fourier transform. The function $\mathcal{C}(\vecv)$ makes the random process effectively non-Gaussian and breaks the property of Gaussian distributions that the even moments are powers of the variance $\sigma^2$. 

A very important result is the anticorrelation of the angular momentum magnitude with the peak height $\nu$, i.e. especially the angular momentum variance decreases if peaks of higher overdensity are selected. \citet{1984Natur.311..517B}, \citet{1986ApJ...301...65H}, \citet{1988ApJ...329....8H} and \citet{1988MNRAS.232..339H} investigate this effect in an analytical derivation as well as by computing a realisation of a Gaussian random field and measuring the tidal shear and inertia at peaks, deriving results on the statistics of the angular momentum and of the spin parameter $\lambda$ as a function of $\nu=\delta/\sigma_0$: The spin of an object depends on the time during which its dynamics is described by linear perturbation theory and during which tidal torquing is effective, and on the correlation of inertia and tidal shear tensors, which depends on the randomness of the field. Typical variations of the angular momentum amount to a decrease of $\sim 30\%$ in amplitude if the peak height is increased from $\nu=1$ to $\nu=2$.

% --- subsection: Angular momenta and peak shape distributions --- %
\subsection{Angular momenta and peak shape distributions}\label{sect_distribution_peak}
\citet{1996MNRAS.282..436C} carry out a computation of the variance $\bra\vecl^2\ket$ resulting from an unconstrained random process, effectively by embedding ellipsoidal haloes with axis ratios drawn from the correct multivariate Gaussian distribution at random positions in the shear flows of the large-scale structure, and generalise this to the derivation of the angular momentum probability distributions. \spirou{They assume no correlation between the tidal shear and the inertia} and start with an evaluation of the tidal shear variance (using statistical isotropy)
\begin{equation}
\bra\Psi_{\beta\sigma}\Psi_{\beta^\prime\sigma^\prime}\ket = 
\left(
\delta_{\beta\sigma}\delta_{\beta^\prime\sigma^\prime} + 
\delta_{\beta\beta^\prime}\delta_{\sigma\sigma^\prime} + 
\delta_{\beta\sigma^\prime}\delta_{\beta^\prime\sigma}
\right)\Phi,
\label{eqn_tidal_shear_isotropy}
\end{equation}
with the abbrevitation
\begin{equation}
\Phi = \frac{4\pi}{15}\int\frac{\dd k}{(2\pi)^3}\: k^6 P_\Psi(k) W^2(kR),
\end{equation}
directly computed from the spectrum $P_\Psi(k)$ of the Zel'dovich-potential $\Psi$, smoothed on the scale $R$. In this way, they can link the angular momentum variance to the shape of the protogalactic object, described by its inertia $I_{\alpha\beta}$,
\begin{equation}
\bra\vecl^2\ket = \frac{2}{15}a^4 \dot{D}_+^2\left(3 I_{\alpha\beta} I_{\alpha\beta} - (I_{\alpha\alpha})^2\right)
\int\frac{\dd k}{2\pi^2}\:k^6P_\Psi(k) W^2(kR). 
\label{eqn_lmomentum_invariant}
\end{equation}
This expression, being scalar and independent of the coordinate system, allows to transform into the principal axis system of the inertia tensor and to simplfy the expression,
\begin{equation}
3 I_{\alpha\beta} I_{\alpha\beta} - (I_{\alpha\alpha})^2 = 2(\mu_1^2-3\mu_2),
\end{equation}
where $\mu_1$ and $\mu_2$ are the invariants of the inertia tensor: $\mu_1 = \trace \matrixi = \iota_1+\iota_2+\iota_3$, and $\mu_2 = \iota_1\iota_2 + \iota_1\iota_3 + \iota_2\iota_3$, formed from the eigenvalues $\iota_i$, $i=1,2,3$, of the inertia tensor. Specifically, the quantities $\mu_1$ and $\mu_2$ are invariant under orthogonal transformations, and in addition there exists a third invariant, which is the determinant $\mu_3=\iota_1\iota_2\iota_3$ of the inertia tensor. A spherical object has 3 identical eigenvalues $\iota_1=\iota_2=\iota_3\equiv\iota$, and therefore $\mu_1 = 3\iota$, $\mu_2 = 3\iota^2$, and hence the expression $\mu_1^2-3\mu_2$ vanishes, in accordance with the fact that spherical protohaloes can not acquire angular momentum by tidal torquing. For a non-spherical volume, $\mu_1^2-3\mu_2>0$, such that tidal torquing can be effective.

\citet{1996MNRAS.282..436C} now link the angular momentum distribution to the distribution of peak shapes by estimating the inertia of a halo forming at a peak from the local curvature of the density field: Clearly, a small value of the local curvature corresponds to a large object, with a large inertia. They show that the term $\mu_1^2-3\mu_2$ replacing the inertia in eqn.~(\ref{eqn_lmomentum_invariant}) depends on the ellipticity $e$ and prolateness $p$ of the protogalactic object,
\begin{equation}
\mu_1^2-3\mu_2 = \frac{2^{11}3^4}{5^2}\pi^2\eta_0^2\left(\frac{\nu}{x}\frac{\sigma_0}{\sigma_2}\right)^5\frac{\mathcal{A}(e,p)}{\mathcal{B}^3(e,p)},
\end{equation}
with $\eta_0=\rho_m a^3$. The ellipticity $e$ and the prolateness $p$ are related to the eigenvalues $\lambda_i$, $i=1,2,3$, of the curvature tensor $\partial_\alpha\partial_\beta\delta$,
\begin{eqnarray}
e & = & \frac{\lambda_1 - \lambda_3}{2(\lambda_1+\lambda_2+\lambda_3)},\\
p & = & \frac{\lambda_1 - 2\lambda_2 + \lambda_3}{2(\lambda_1+\lambda_2+\lambda_3)}.
\end{eqnarray}
For a triaxial ellipsoid, the parameter $e\geq0$ measures the ellipticity, and $p$ gives information whether the object is oblate ($0\leq p\leq e$) or prolate ($-e\leq p\leq0$). For an alternative description of the local curvature in terms of $e$ and $p$ instead of the eigenvalues $\lambda_i$, $i=1,2,3$, the additional parameter $x$,
\begin{equation}
x = \frac{\lambda_1+\lambda_2+\lambda_3}{\sigma_2}
\end{equation}
is needed. The moments $\sigma_n(R)$ derived from the power spectrum of the smoothed potential can be computed using
\begin{equation}
\sigma_n^2(R) = \int\frac{\dd k}{2\pi^2}\:k^{6+2n}P_\Psi(k) W^2(kR).
\label{eqn_sigman_moment}
\end{equation}
The polynomials $\mathcal{A}$ and $\mathcal{B}$ are given by:
\begin{eqnarray}
\mathcal{A}(e,p) & = & \left[p(1+p)\right]^2 + 3e^2(1-6p+2p^2+3e^2),\\
\mathcal{B}(e,p) & = & (1-2p)\left[(1+p)^2-9e^2\right].
\end{eqnarray}
Using these relations, the final step is to relate the angular momentum distribution to the probability distribution of peak shapes $p(\nu,x,e,p)\dd\nu\:\dd x\:\dd e\:\dd p$ derived in \citet{1986ApJ...304...15B},
\begin{equation}
p(\nu,x,e,p)\propto x^8\mathcal{W}(e,p)\exp\left(-\frac{\nu^2}{2}-\frac{5}{2}x^2(3e^2+p^2)-\frac{(x-\gamma\nu)^2}{2(1-\gamma^2)}\right)
\end{equation}
with the function $\mathcal{W}(e,p) = e(e^2-p^2)\mathcal{B}(e,p)$. The probability distribution is defined in the allowed triangular domain $0\leq e\leq1/4$, $-e\leq p\leq e$ and $1/4\leq e\leq 1/2$, $3e-1\leq p\leq e$. In a quantitative investigation, \citet{1996MNRAS.282..436C} find that by choosing typical peak heights, the value of the specific angular momentum of spiral galaxies can be reproduced, but they find discrepancies for higher peaks corresponding to elliptical galaxies, as the specific angular momentum predicted is too large. Another important result that the median of derived distribution of the spin parameter $\lambda$ from this model agrees well with the value observed in spiral galaxies.

% --- subsection: angular momentum biasing --- %
\subsection{Angular momentum biasing}\label{sect_distribution_bias}
The restriction of the random process for angular momenta to peak locations with has two consequences, which effectively introduce a biased angular momentum distribution. Firstly, as shown by \citet{1988MNRAS.232..339H}, who perform a direct computation of the distribution $p(L)\dd L$ by setting up the joint Gaussian random process outlined above and integrating out the peak restriction in the derivation of $p(L)\dd L$, the angular momentum distribution is non-Gaussian and although the mean angular momentum needs necessarily to be zero due to isotropy (a nonzero mean angular momentum of the large-scale structure would define a preferential direction), the odd higher order moments of the angular momentum distribution have no reason to vanish. Additionally, they provide fitting formul{\ae} for the resulting angular momentum distribution $p(L)\dd L$.

The second consequence of the peak restriction is a reduced variance of the angular momentum field, compared to that of a continuous field, due to the covariance of the curvature of the density field with the tidal shear. The magnitude of this effect is determined by \citet{1996MNRAS.282..436C}, who show that the restriction to peak location reduces the variance of the tidal shear by $(1-\gamma^2)$,
\begin{equation}
\bra\Psi_{\beta\sigma}\Psi_{\beta^\prime\sigma^\prime}\ket_\mathrm{peak} = (1-\gamma^2)\: 
\bra\Psi_{\beta\sigma}\Psi_{\beta^\prime\sigma^\prime}\ket,
\end{equation}
with the value $\gamma$,
\begin{equation}
\gamma = \frac{\sigma_1^2}{\sigma_0\sigma_2}.
\end{equation}
This reduction, which is mass-scale dependent due to the smoothing scale $R$, is due to the fact that the random field is constrained to have a peak with a specified shape, and is restricted in its fluctuations. $\sigma_n$ are given by eqn.~(\ref{eqn_sigman_moment}).

The idea of the derivation outlined in Sect.~\ref{sect_distribution_peak} was to place protohaloes with axis ratios from the multivariate distribution $p(\nu,x,e,p)$ at random positions in the large-scale structure and to torque them up with the local tidal fields, such that one obtains the probability distribution of the angular momentum. This procedure assumes the validity of two simplifications: The fluctuation amplitude of the tidal shear field is the same restricted to peak positions and in the case of a continuous field, as discussed above. Secondly, there are no correlations between the inertia of the protohaloes which form at peak positions and the tidal shear field. This is seemingly inconsistent, as the shape of the potential and hence the density field needs to be compatible with the shape of the object which is forming at a particular position - and one would need to take account of the correlations between the inertia and tidal shear tensors. In the case of a Gaussian random field, however, it is possible to transform in the eigenframe of the inertia tensor, such that only the off-diagonal elements of the tidal shear tensor are needed for the computation of the angular momentum, due to the occurrence of the Levi-Civit\`{a}-symbol in eqn.~(\ref{eqn_l_definition}). These elements turn out to be statistically uncorrelated with the diagonal entries of the inertia tensor, such that the assumption of uncorrelated tidal shear and inertia distributions is justified, the physical reason of this being again the Poisson equation, which relates the density field to the trace of the tidal shear.

In particular, this fortunate coincidence can not be used in the case of angular momentum correlations, for which the full correlations between the inertia and shear across two points in the large-scale structure as well as between themselves at the same location has to be taken account of, as will be demonstrated in Sect.~\ref{sect_correlation_basic}.

% --- subsection: nonlinear torquing --- %
\subsection{Nonlinear tidal torquing}\label{sect_distribution_nonlinear}
The theory of angular momentum acquisition by tidal shearing has been extended to nonlinear stages by using second order perturbation theory \citep{1996MNRAS.282..455C} and to include effects of non-Gaussian initial perturbations \citep{1997MNRAS.292..225C}. \citet{1996MNRAS.282..455C} present a perturbative analysis which uses higher-order correction terms in Lagrangian perturbation theory. They find that the variance $\bra\vecl^2\ket$ of the angular momentum is increases by a factor of 1.6 compared to the first-order computation, or equivalently, that the angular momenta are higher by about 30\%, and emphasise that the initial torque is a good estimate for the total torque acting on the protogalactic object. The second paper \citep{1997MNRAS.292..225C} investigates the sensitivity of the angular momentum statistics on non-Gaussian initial conditions, using the $\chi^2$-model and a model with a log-normal distribution of the density amplitudes and following tidal torquing in second-order Lagrangian perturbation theory. Their derivation is generic such that it can be applied to both primordial non-Gaussianities in the density field or to Gaussianities generated by nonlinear structure formation. They find that angular momenta grow from non-Gaussian perturbations at a lower rate ($\propto t^{8/3}$) compared to Gaussian perturbations ($\propto t^{10/3}$) in an SCDM-cosmology with $\Omega_m\equiv1$. The absolute magnitude of the correctional terms at turn-around can be larger that the linear estimates, however, depending on the total skewness.

% --- subsection: spin parameter --- %
\subsection{Spin parameter of galaxies and galaxy formation}\label{sect_distribution_spin}
Not too surprisingly, the halo angular momentum has far-reaching consequences for the internal structure of the galaxy which is hosted by a dark matter structure. This is commonly expressed by the dimensionless spin parameter $\lambda$ introduced by \citet{1969ApJ...155..393P},
\begin{equation}
\lambda = \frac{L\sqrt{E}}{G M^{5/2}},
\label{eqn_lambda_def}
\end{equation}
for a halo with angular momentum $L$, gravitational binding energy $E$ and mass $M$. $G$ denotes Newton's gravitational constant. $\lambda$  corresponds to the ratio between the observed angular velocity of a galaxy $\omega$ and the angular velocity needed for rotational support $\omega_0$:
\begin{equation}
\lambda = \frac{\omega}{\omega_0} = \frac{L/(MR^2)}{\sqrt{GM/R^3}},
\end{equation}
and numerical simulations suggest that the scale radius of exponential disks is in fact $\propto\lambda$. $\lambda$ depends on the mass scale of the dark matter halo, assuming values of up to $\lambda=0.5$ for spirals, and values of $\lambda\simeq0.05$ for elliptical galaxies, with a very large dispersion \citep{1979MNRAS.186..133E, 1987ApJ...319..575B}, suggesting that dark matter haloes are stabilised mainly by anisotropic velocity dispersion. 

The details of the angular momentum profile inside a dark matter halo is investigated in detail by \citet{2001ApJ...555..240B}, who report that the cumulative angular momentum profile $M(<j)$ of haloes forming in an $n$-body simulation is fitted by a function of the type
\begin{equation}
M(<j) \propto \frac{\mu j}{j+j_0},
\end{equation}
where $j=L/M$ is the specific angular momentum. The profile has an implicit maximum value of $j_\mathrm{max}=j_0/(\mu-1)$, is of approximate power-law shape for $j\lsim j_0$ anf flattens out for $j\gsim j_0$. The two parameters $\mu$ and $j_0$ are adapted to fit experimental data. The angular momentum profile clearly deviate from solid-body rotation, implying that dark matter haloes exhibit angular momentum structure. The averaged angular momentum directions in shells, however, were found to be very well aligned with each other.

Analytical derivations \citep{1969ApJ...155..393P, 1988MNRAS.232..339H, 1995MNRAS.272..570S, 1996MNRAS.282..436C} as well as numerical simulations \citep{1992ApJ...399..405W, 2002ApJ...581..799V, 2007MNRAS.380L..58D} using realistic CDM transfer functions for the initial density field find $\lambda$ to be approximately log-normal distributed,
\begin{equation}
p(\lambda)\dd\lambda = \frac{1}{\sqrt{2\pi}\sigma_\lambda}\exp\left(-\frac{\ln^2(\lambda/\mu_\lambda)}{2\sigma_\lambda^2}\right)\frac{\dd\lambda}{\lambda}
\end{equation}
with the parameters $0.03\leq\mu_\lambda\leq 0.05$ and $0.5\leq\sigma_\lambda\leq 0.7$. The mean of the distribution is related to $\mu_\lambda$ by $\bra\lambda\ket\simeq1.078\mu_\lambda$ for $\sigma_\lambda=0.6$. 

Using the results from $n$-body simulations, \citet{1998MNRAS.295..319M} and \citet{1998ApJ...505...37A} consider the formation of galactic disks inside dark matter haloes, while assuming that the baryonic component is torqued up to the same specific angular momentum as the dark matter, and forms a stable, rotationally supported disk with exponential surface density profile, whose mass makes up a few percent of the host halo. In this way they are able to reproduce the relation between disk size and rotation velocity, as well as the slope and scatter of the Tully-Fisher relation (i.e. the relation between rotational velocity and luminosity), calibrating the zero-point of the Tully-Fisher relation by fixing the mass to light ratio. Further results include rotation curves for galactic disks embedded into haloes for a given spin parameter $\lambda$, which appear very convincing. The interplay between dark matter and baryonic dynamics is analysed in the papers by \citet{2006ApJ...638L..13T} and \citet{2006ApJ...649..591T}, who find higher spin parameters including baryonic physics compared to the dark matter case and a flattening of the central halo density profile.

Other interesting results include the paper by \citet{1995ApJ...443...11E}, who suggest that quasar progenitors form by gravitational collapse of perturbations with low specific angular momentum on the mass scale of $10^{5}M_\odot/h$. Inside these dark matter haloes, they consider the formation of a galactic disk on the viscous time scale and conclude that the number density of these objects is $\simeq10^{-3}/(\mathrm{Mpc}/h)^3$, in accordance with the spatial density of quasars. 

\citet{2002ApJ...576...21V} investigate the spin-parameter distribution of the baryonic and the dark matter components and find that they are very similar, suggesting that angular momentum is conserved during disk formation. While the distributions are similar, the angular momentum directions were found to be different by a median misalignment angle as large as $30^\circ$. \spirou{In a careful analysis of numerical data they conclude that in the course of galaxy formation $(i)$ the specific angular momentum of the galactic disk results from tidal torquing, $(ii)$ the specific angular momentum of the baryonic component is equal to that of the dark matter, but that $(iii)$ the specific angular momentum is not conserved during disk formation, due to virialisation processes in which the baryonic angular momentum gets redistributed, while different relaxation mechanisms are accessible to the baryonic component and to the dark matter.}

\citet{2005ApJ...627L..17B} analyse a set of hydrodynamical simulations of galaxy formation and find that the disk has a minor influence on the overall halo structure, and that the disk is aligned with the host halo inside $0.1 r_\mathrm{vir}$, which is a reassuring result for the usage of angular-momentum based ellipticity correlation models in gravitational lensing. \spirou{The review by \citet{2008arXiv0801.3845M} on computer simulations of galaxy formation lists some of the numerical issues related to angular momenta, e.g. the fact that angular momentum is lost artificially by the simulation codes. In addition, they report that the angular momentum of a simulated galaxy is resolution dependent: Increasing the number of particles by a factor of 400 yields angular momenta larger by 30\%, emphasising the difficulty of numerical simulations of galaxy formation, especially regarding the physics of the interstellar medium. The reliability of the codes for determining the alignment of the baryons inside the dark matter host structure is still an open question.}

\spirou{The measurement of the spin parameter from observational data is the topic of the papers by \citet{2007MNRAS.375..163H} and by \citet{2008arXiv0802.1934B}, the major issue being that none of the quantities in the definition of $\lambda$ (c.f. eqn.~\ref{eqn_lambda_def}) is an observable. Both papers use SDSS-data and infer $\lambda$ by constructing a model of the galaxy consisting of an exponential disk placed in a truncated isothermal sphere, which ensures a flat rotation curve of the disk due to $\rho_\mathrm{halo}(r)\propto r^{-2}$. In this model, the halo dominates $L$, $E$ and $M$ and the disk is a mere tracer of the dynamics. By assuming that the specific angular momenta of the halo and of the disk are equal, one can infer $\lambda$ from an observation of the rotational velocity of the disk, if in addition the validity of a Tully-Fisher relation between the disk's rotational velocity and mass is assumed, and if a constant factor relates the disk mass to the halo mass. Both papers report that the measured distribution $p(\lambda)\dd\lambda$ agrees well with log-normal distribution suggested by numerical data, recovering values of $\bra\lambda\ket\simeq0.05$. \citet{2008arXiv0802.1934B} in addition investigate the influence of spin onto the star formation and find an anticorrelation between spin and stellar mass, as well as the fact that galaxies with recent star formation tend to have higher spins. They explain these findings in terms of a thinner disk of systems with high values of $\lambda$, which might be less efficient in forming stars, leaving a larger reservoir of gas today.}

% --- section: angular momentum correlations ---%
\section{angular momentum correlation}\label{sect_correlation}
This section treats the correlation in angular momentum between neighbouring haloes: The formalism of the previous section is extended for computation of the 2-point variance of the angular momentum field and hence the correlation function (Sect.~\ref{sect_correlation_basic}). An important simplification, in which the correlations between the inertia tensors is disentangled from those of the tidal shear is discussed, (Sect.~\ref{sect_correlation_decomposition}), followed by outlining a very practical model due to \citet{2000ApJ...532L...5L} which links the angular momentum directly to the tidal shear (Sect.~\ref{sect_correlation_parameterisation}) by using a parameter to allow for skewing between the shear and inertia eigensystems.

% --- subsection: ansatz for the angular momentum correlations --- %
\subsection{Ansatz for the angular momentum correlations}\label{sect_correlation_basic}
Generalisation of the relations in Sect.~\ref{sect_distribution_moments} to include a second peak results in the correlation function $\bra L_\alpha(\vecx) L_{\alpha^\prime}(\vecx^\prime)\ket$ of the angular momenta, with the Gaussian probability density $p(\vecw)\dd\vecw=p(\vecv,\vecv^\prime)\dd\vecv\dd\vecv^\prime$:
\begin{equation}
\bra L_\alpha(\vecx) L_{\alpha^\prime}(\vecx^\prime)\ket = 
\label{eqn_ll_corr}
\end{equation}
\begin{displaymath}
\quad\frac{1}{n_\mathrm{peak}^2(>\nu)}
\int\dd\vecv\mathcal{C}(\vecv)\:\int\dd\vecv^\prime\mathcal{C}(\vecv^\prime)\: 
L_\alpha(\vecv) L_{\alpha^\prime}(\vecv^\prime) p(\vecv,\vecv^\prime).
\end{displaymath}
In general, the thresholds $\nu$, $\nu^\prime$ imposed on the peaks are equal  \citep[see ][for a discussion of the additional complication of unequal peak heights]{2005ApJ...632..706L}. As derived in Sect.~\ref{sect_torquing_zeldovich}, the angular momentum $L_\alpha$ depends on the product the inertia tensor $I_{\beta\sigma}$ and the tidal shear $\Psi_{\sigma\gamma}$:
\begin{equation}
L_\alpha 
= a^2 \dot{D}_+ \epsilon_{\alpha\beta\gamma}\sum_\sigma I_{\beta\sigma} \Psi_{\sigma\gamma}
= a^2 \dot{D}_+ \epsilon_{\alpha\beta\gamma}X_{\beta\gamma}, 
\end{equation}
if the acquisition of angular momentum of a protogalactic object is described in the Zel'dovich approximation. Then, the correlation of the angular momentum components becomes:
\begin{equation}
\bra L_\alpha(\vecx) L_{\alpha^\prime}(\vecx^\prime)\ket = a^4 \dot{D}_+^2
\epsilon_{\alpha\beta\gamma}\epsilon_{\alpha^\prime\beta^\prime\gamma^\prime}
\bra X_{\beta\gamma}(\vecx) X_{\beta^\prime\gamma^\prime}(\vecx^\prime)\ket.
\label{eqn_derivation_ll_corr_matrix}
\end{equation}
In the next step I replace the 1d variance of the components $L_\alpha$ in the correlation function by the 3d variance of the full vector $\vecl$ by taking the trace of eqn.~(\ref{eqn_derivation_ll_corr_matrix}),
\begin{equation}
C_L(r)
\equiv\trace\bra L_\alpha(\vecx) L_{\alpha^\prime}(\vecx^\prime)\ket
= \bra L_\alpha(\vecx) L_{\alpha}(\vecx^\prime)\ket,
\label{eqn_trace_ll_corr}
\end{equation}
which has the advantage of being a coordinate-frame independent quantity and allows the usage of the relation $\epsilon_{\alpha\beta\gamma}\epsilon_{\alpha\beta^\prime\gamma^\prime} = \delta_{\beta\beta^\prime}\delta_{\gamma\gamma^\prime} - \delta_{\beta\gamma^\prime}\delta_{\beta^\prime\gamma}$ for reducing the product of the two $\epsilon_{\alpha\beta\gamma}$-symbols:
\begin{equation}
\bra L_\alpha(\vecx) L_\alpha(\vecx^\prime)\ket = a^4 \dot{D}_+^2
\left[
\bra X_{\beta\gamma}(\vecx^\prime) X_{\beta\gamma}(\vecx) \ket - \bra X_{\beta\gamma}(\vecx) X_{\gamma\beta}(\vecx^\prime)\ket
\right],
\end{equation}
where the order of the indices in the last term is interchanged. In matrix notation, the correlation function $C_L(r)$ reads:
\begin{equation}
C_L(r)
= \trace\bra \vecl(\vecx) \vecl^t(\vecx^\prime)\ket 
= a^4 \dot{D}_+^2 \:\trace
\left[\bra\matrixx(\vecx^\prime)\matrixx(\vecx)\ket - \bra\matrixx(\vecx)\matrixx^t(\vecx^\prime)\ket\right],
\label{eqn_derivation_ll_ttranspose}
\end{equation}
which is non-vanishing for general asymmetric matrices $X_{\beta\gamma}$ due to the matrix transposition in the last term. In the correlation function of the angular momenta $C_L(r)=\trace\bra\vecl^t(\vecx)\vecl(\vecx^\prime)\ket$ (c.f. eqn.~\ref{eqn_derivation_ll_ttranspose}), the dependence of $\vecl$ on the commutator $\matrixx^-$ translates into the asymmetric quadratic form $\bra\matrixx(\vecx)\matrixx(\vecx^\prime)-\matrixx(\vecx)\matrixx^t(\vecx^\prime)\ket$ with the matrix transpose in the second term carrying the signal: A common eigensystem of the inertia and shear tensors, being both symmetric matrices, would have the consequence that $\matrixx$ would be a symmetric matrix, $\matrixx=\matrixx^+=\matrixx^t$, and the correlation function $C_L(r)$ would vanish. 

In contrast, the signal is maximised, if the shear and inertia eigensystems are unaligned, i.e. if $\matrixx$ is purely antisymmetric, $\matrixx=\matrixx^-$ and $\matrixx^+=0$. In that case $\matrixx^t=(\matrixx^-)^t=-\matrixx^-=-\matrixx$ and the angular momentum auto-correlation function $C_L^\mathrm{max}(r)$ is truly quadratic:
\begin{equation}
C_L^\mathrm{max}(r) = 2a^4 \dot{D}_+^2 \:\trace
\left[\bra\matrixx(\vecx^\prime)\matrixx(\vecx)\ket\right],
\label{eqn_ll_quadratic}
\end{equation}
which illustrates the effect of partial alignment of the shear and inertia eigensystems, reducing the variance compared to the case where the shear and inertia eigensystems are maximally misaligned. Additionally, eqn.~(\ref{eqn_derivation_ll_ttranspose}) yields negative values for antiparallel alignment of the angular momenta of neighbouring haloes \citep[see][together with a discussion of the symmetry properties of eqns.~\ref{eqn_derivation_ll_ttranspose} and~\ref{eqn_ll_quadratic}]{my_paper}.

% --- subsection: decomposition of tidal shear and inertia correlations --- %
\subsection{Decomposition of tidal shear and inertia correlations}\label{sect_correlation_decomposition}
As already pointed out, the correlations in the angular momentum field result from an interplay of the long-ranged correlations in the tidal shear field (which has the same correlation length as the density field) and the much shorter range correlation of the inertia (which is related to the curvature of the density field, i.e. its second derivative). This means that at large separation distances the random processes determining the inertias of the two haloes are independent and only take account of the local shape of the density field, encoded in the tidal shear. Motivated by this observation, \citet{1996MNRAS.282..436C} propose the decomposition
\begin{equation}
\bra\matrixi(\vecx)\bmath{\Psi}(\vecx)\:\matrixi(\vecx^\prime)\bmath{\Psi}(\vecx^\prime)\ket =
P(\matrixi|\bmath{\Psi}(\vecx))\: P(\matrixi|\bmath{\Psi}(\vecx^\prime))\:
\bra\bmath{\Psi}(\vecx)\:\bmath{\Psi}(\vecx^\prime)\ket,
\label{eqn_decomposition}
\end{equation}
with two tensor-valued conditional probability densities $P(\matrixi|\bmath{\Psi}(\vecx))=\bra\matrixi\:\bmath{\Psi}(\vecx)\ket$ which give the distribution of obtaining a halo with a certain inertia given a particular shape of the density field, or equivalently, tidal shear field. This approach is applicable on separations larger than the corelation length of the inertia field and greatly reduces the complexity of the evaluation of the 4-point object $\bra\matrixi(\vecx)\bmath{\Psi}(\vecx)\:\matrixi(\vecx^\prime)\bmath{\Psi}(\vecx^\prime)\ket$. 

Treating $\Psi_{\sigma\gamma}$ as a continuous field, \citet{2001MNRAS.323..713C} compute the correlation function $\bra\Psi_{\sigma\gamma}\Psi_{\sigma^\prime\gamma^\prime}\ket$ of the tidal shear field, which has to reflect the symmetry relations of the tidal shear tensor (by carrying out a decomposition into scalar, vectorial and tensorial components), as well as to distinguish between longitudinal and transverse correlations with respect to the separation vector.

% --- subsection: parameterised angular momentum correlations --- %
\subsection{Parameterised angular momentum correlations}\label{sect_correlation_parameterisation}
\citet{2000ApJ...532L...5L} and \citet{2001ApJ...559..552C} proposed to link the angular momentum directly to the tidal shear and to use an effective parameterisation of the misalignment between inertia and shear:
\begin{equation}
\bra L_\alpha L_{\alpha^\prime}\ket = 
\frac{\bra\vecl^2\ket}{3}
\left(\frac{1+a}{3}\delta_{\alpha\alpha^\prime} - a\sum_\gamma \hat{\Psi}_{\alpha\gamma}\hat{\Psi}_{\gamma\alpha^\prime}\right),
\label{eqn_ll_crittenden}
\end{equation}
where $\hat{\Psi}$ is the unit-normalised ($\hat{\Psi}_{\alpha\beta}\hat{\Psi}_{\alpha\beta}=1$) traceless ($\mathrm{tr}\hat{\Psi} = 0$) tidal shear tensor, which can easily be derived using
\begin{equation}
\bar{\Psi}_{\alpha\beta} = \Psi_{\alpha\beta} - \frac{\trace\Psi}{3}\delta_{\alpha\beta},
\end{equation}
and rescaling $\hat{\Psi} = \bar{\Psi}/\left|\bar{\Psi}\right|$. The derivation of eqn.~(\ref{eqn_ll_crittenden}) assumes no correlation between the inertia and shear tensors, as explained in the previous section, uses the decomposition eqn.~(\ref{eqn_decomposition}) applicable at large separations and assumes statistical isotropy of the inertia tensor, i.e. an evaluation of the inertia correlations in analogy to eqn.~(\ref{eqn_tidal_shear_isotropy}),
\begin{equation}
\bra I_{\beta\sigma} I_{\beta^\prime\sigma^\prime}\ket = 
\frac{1}{3}
\left(
\delta_{\beta\sigma}\delta_{\beta^\prime\sigma^\prime} + 
\delta_{\beta\beta^\prime}\delta_{\sigma\sigma^\prime} + 
\delta_{\beta\sigma^\prime}\delta_{\beta^\prime\sigma}
\right)
\bra I_{11}^2 + I_{22}^2 + I_{33}^2\ket.
\end{equation}
The parameter $a$, which measures the degree of misalignment between the shear and inertia eigensystems assumes values between $a=0$ corresponding to the case of perfect correlation between inertia and shear, and the maximum value of $a=3/5$, if inertia and shear are mutually uncorrelated and if there are no nonlinear contributions to the angular momentum. If $a=0$, one recovers the isotropy relation $\bra L_\alpha L_{\alpha^\prime}\ket = \delta_{\alpha\alpha^\prime}/3$, and quite generally, a stronger correlation between the shear and inertia makes the angular momentum more random. In numerical simuations, $a$ was measured to assume a value of $a=0.24$ \citep{2000ApJ...532L...5L}. 

\spirou{A subtle point is the parameterisation $a$ in eqn.~(\ref{eqn_ll_crittenden}): In the derivation of the formula, \citet{2000ApJ...532L...5L} assumed uncorrelated shear and inertia tensors, as stated above, and proceed by calculating the variance of the angular momentum for a given shear, for which they find a term quadratic in the tidal shear as a correction to the variance resulting from random orientation of the angular momenta, because haloes tend to align their angular momentum with the intermediate eigenaxis of the shear tensor. They argue that this dependence on shear should vary between 0 and 1, corresponding to the cases of random orientation of angular momenta, and of angular momenta fixed by the shear eigenframe, respectively. By setting up a Gaussian random process and by calculating the variance of the angular momentum direction for a given tidal shear in the limit of weak alignment, they recover eqn.~(\ref{eqn_ll_crittenden}), which can be scaled to the variance of the full angular momentum by the expectation value $\bra\vecl^2\ket/3$.}

\citet{2001ApJ...555..106L} continue by defining their angular momentum correlation functions $\eta(r)$ and $\eta_2(r)$ as the 2-point variance of the angular momentum directions $\hat{L}=\vecl/L$,
\begin{equation}
\eta_2(r) = 
\langle\left|\hat{L}(\vecx)\hat{L}(\vecx+\vecr)\right|\rangle - \frac{1}{3}
\simeq \frac{a^2}{6}\frac{\xi_R^2(r)}{\xi_R^2(0)},
\end{equation}
and
\begin{equation}
\eta(r) = \langle\hat{L}(\vecx)\hat{L}(\vecx+\vecr)\rangle,
\end{equation}
with the filtered correlation function $\xi_R(r)$ of the density field,
\begin{equation}
\xi_R(r) = \int\frac{\dd k}{2\pi^2}\: k^2 P(k) W^2(kR) j_0(kr).
\end{equation}
The proportionality to the square $\xi^2_R(r)$ is a direct consequence of the parameterisation eqn.~(\ref{eqn_ll_crittenden}) with the squared tidal shear, and the term $1/3$ makes the correlation function $\eta_2(r)$ vanish in the case of no alignment. The length scale $R$ is related to the mass scale $M$ by the usual relation $M=4\pi\Omega_m\rho_\mathrm{crit}R^3/3$, applicable for a top-hat filter kernel.

Here, it should be emphasised that eqn.~(\ref{eqn_ll_crittenden}) provides a correlation rather between the angular momentum axes than between the angular momenta themselves as it is unable to provide the full directional information on the angular momentum vector $\vecl$ and can not distinguish between parallel and antiparallel aligned angular momentum vectors, although it is invariant under a global parity transformation. Furthermore, the parameters $a$ and $\bra\vecl^2\ket$ have been measured from collapsed objects in numerical simulations, for which the tidal shear mechanism in its simplest form is not applicable as there are contributions from nonlinear dynamics which exceed the description using the Zel'dovich approximation. Furthermore, it is not straightforward in a numerical simulation to identify and to measure the initial conditions from which an object has formed. Additionally, eqn.~(\ref{eqn_ll_crittenden}) links the correlation properties of the continuous tidal shear field to those of the discrete angular momentum field, without the biasing of angular momentum. Nevertheless, the formula has enormous merits in describing correlations in the angular momentum direction, and is successfully used in intrinsic shape alignment studies, as the galaxy ellipticity does not depend on the magnitude of $\vecl$ and is invariant under the transformation $\vecl\rightarrow -\vecl$.

% --- section: numerical simulations ---%
\section{numerical simulations}\label{sect_simulation}
Angular momentum build-up by tidal torquing and the resulting angular momentum correlations have been extensively tested in numerical $n$-body simulations of cosmic structure formation (Sect.~\ref{sect_simulation_tests}), as well as alignments of angular momenta with ambient cosmological structures such as filaments, sheets and voids (Sect.~\ref{sect_simulation_alignments}). In addition, there are studies that focus on competing mechanisms for angular momentum build-up such as accretion (Sect.~\ref{sect_simulation_accretion}).

% --- subsection: numerical tests of tidal torque theory --- %
\subsection{Numerical tests of tidal torque theory}\label{sect_simulation_tests}
One of the first numerical studies of angular momentum build-up of objects in the large-scale structure was carried out by \citet{1979MNRAS.186..133E}, who ran a simulation of $10^3$ particles, finding a few tens of objects. They were able to reproduce the $L\propto M^{5/3}$-relationship predicted by linear tidal torquing, and analysed the distributions of the spin parameter $\lambda$, recovering the low values for $\lambda$ predicted analytically. Later papers, e.g. \citet{1987ApJ...319..575B}, improve conclusions tremendously due to the increase in number of particles, realistic initial conditions and algorithmic advancement. Their results include an accurate determination of the $L$-$M$-relation, where they find best fit values close to the fiducial $L\propto M^{5/3}$-scaling, as well as  the time evolution of the angular momentum acquisition, with its truncation at turn-around. The distributions of $\lambda$ were found to be insensitive on the particular cosmology and on the slope $n_s$ of the matter power spectrum $P(k)\propto k^{n_s}$. Additionally, they investigated the alignment of the spin axis with the ellipticity axes of the dark matter halo and found the spin to point preferentally perpendicular relative to the major axis of the ellipsoid, before they discuss the relation between $\lambda$ and morphological type, emphasising the importance of baryonic physics. \citet{2003astro.ph.12547P} provides a summary on the relations between halo angular momentum and disk formation, emphasising open questions on the matter.

\citet{2000MNRAS.311..762S} carry out an extensive comparison between the analytical predictions from tidal torquing with numerical $n$-body simulations, confirming the basic tidal torquing mechanism. They find a linear growth with cosmic time until turn-around, and continues quasi-linearly until shell crossing, when the Zel'dovich-approximation ceases to be applicable. Linear tidal torquing was found to overestimate the true turn around by a factor of three (indicating the importance of tidal forces on object formation by collapse) as well as yielding too large values for the total angular momentum. In measuring the mass and binding energy of the dark matter haloes, they confirm the anti-correlation of the spin parameter $\lambda$ with peak height $\nu$ and the angular momentum scaling with mass, $L\propto M^{5/3}$.

Details of tidal torquing as a mechanism of angular momentum build-up are scrutinised in the series of papers written by \citet{2002MNRAS.332..325P} and \citet{2002MNRAS.332..339P}. In the first paper, they compare predictions from tidal torquing to the spin amplitudes of haloes forming in a cosmological $n$-body simulation, assuming that torquing is efficient up to $0.6$ halo's turn around time, and find that the order of magnitude of the angular momentum can be well predicted down to redshifts below $z\simeq3$. Nonlinear influences cause a significant deviation of the angular momentum direction from the direction predicted by tidal torquing, amounting to a mean error of about $50^\circ$, which weakens the spin correlations, and casts doubts on the usefulness of tidal torquing for predicting ellipticity alignments. In particular, \citet{2002MNRAS.332..325P} confirm that the mechanism of tidal torquing in the linear regime is a very good model, and that the truely nonlinear effects at late times cause deviations: Not only is the rotation axis affected, but the amplitude of the angular momentum grows by more than a factor of two between redshifts of $z=3$ and $z=0$ in addition to the prediction of linear tidal torquing. Further results include the angular momentum time evolution in units of the turn-around time $t_0$, which levels out from a linear increase at $\simeq0.6t_0$, in accordance with the tidal torquing model, as well as the distribution of ellipticity $e$ and prolateness $p$, which agrees well with the expected distribution for a Gaussian random field. 

\begin{figure*}
\begin{tabular}{cc}
\resizebox{8cm}{!}{\includegraphics{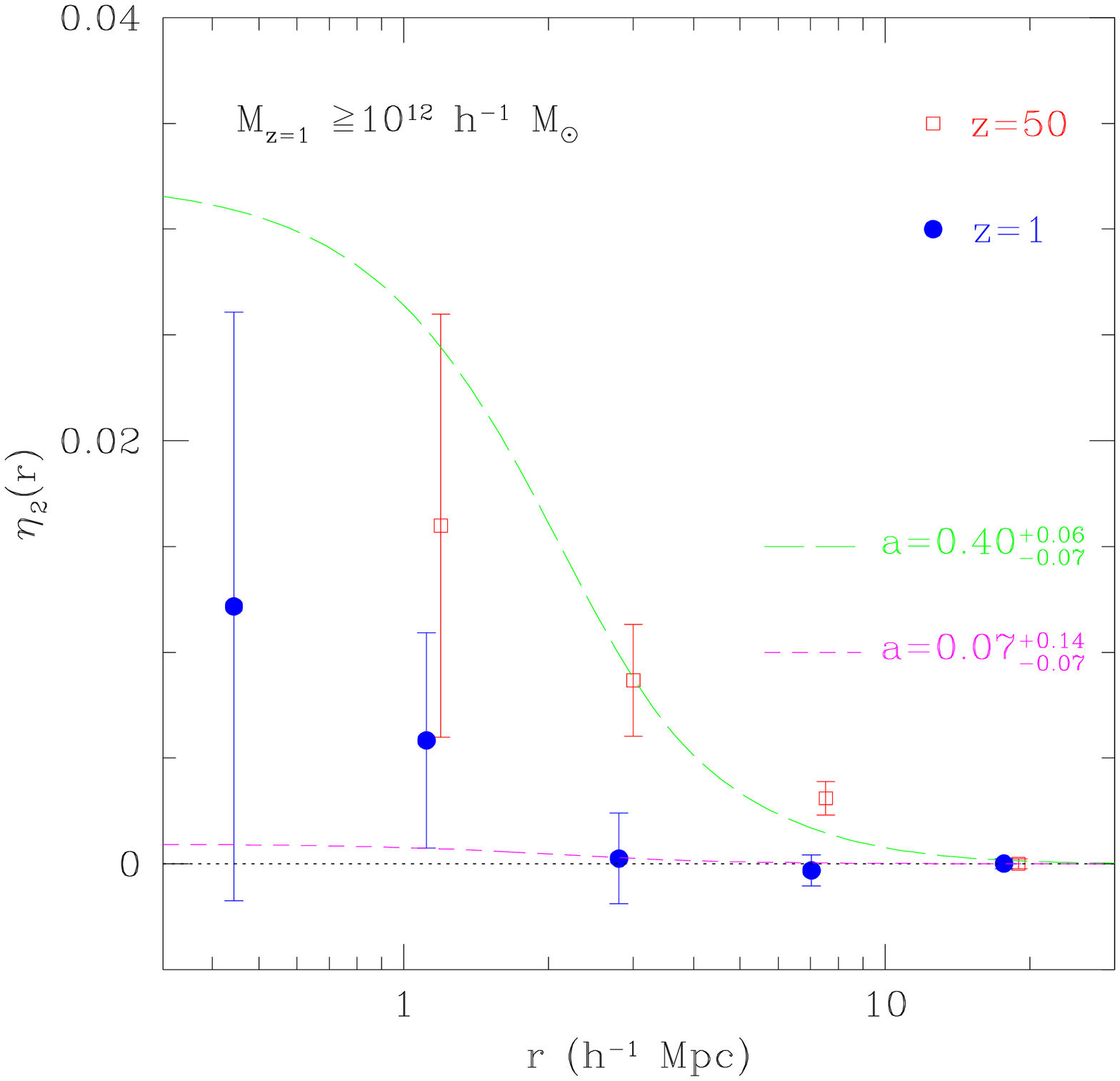}} & 
\resizebox{8cm}{!}{\includegraphics{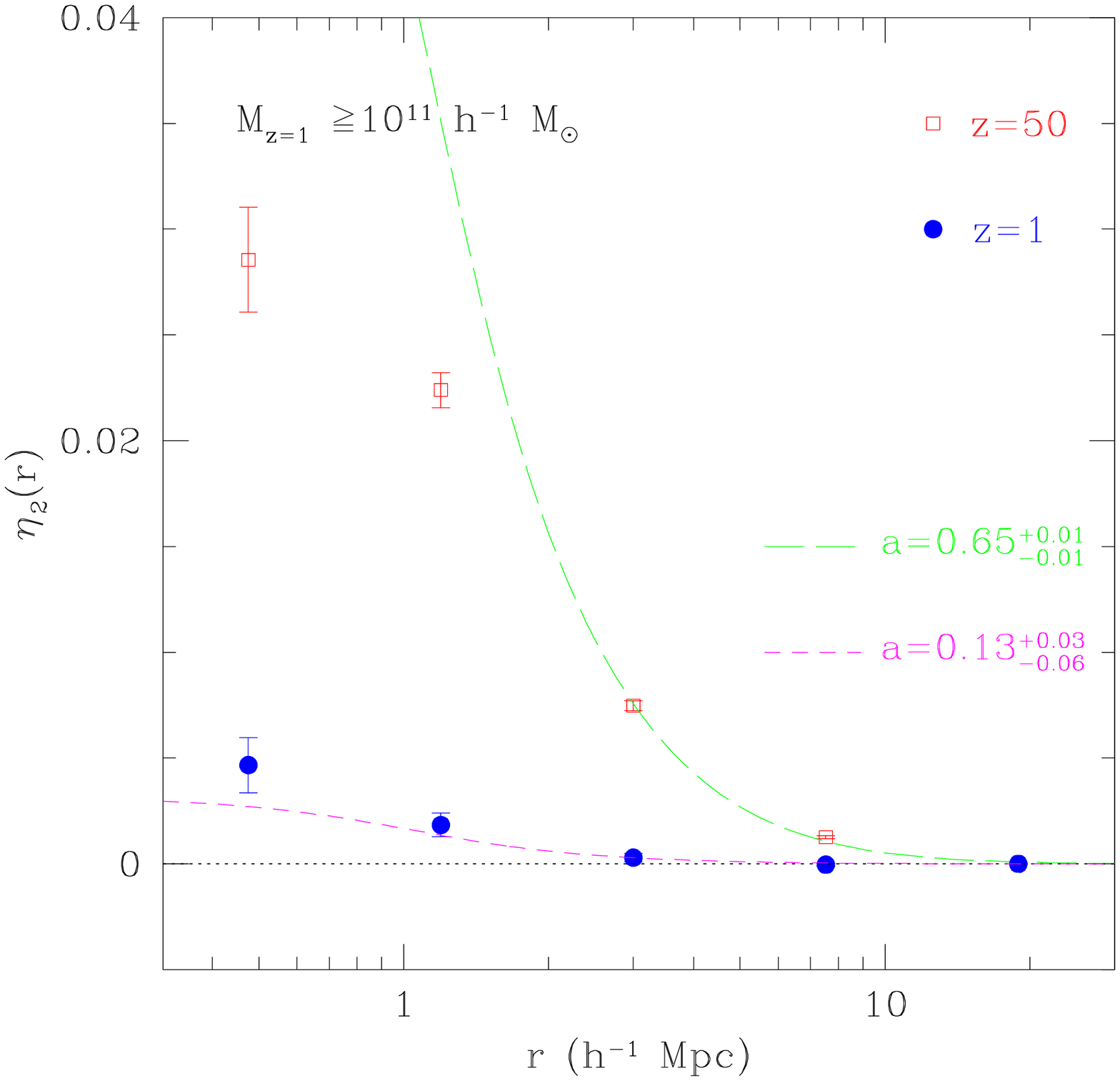}}
\end{tabular}
\caption{Correlation function and $\eta_2(r)$ of the angular momentum directions of low high mass (left) and low mass (right) haloes, comparing the predictions from tidal torquing (open squares) with the results from $n$-body simulations (filled circles). The error bars reflect sampling errors only and ignore the error associated with the spin measurement. The dashed line indicates the theoretical expectation. {\em The figure was kindly provided by C. Porciani and appears in \citet{2002MNRAS.332..339P}. Reproduced with permission from Blackwell Publishing.}}
\label{fig_ll_correlation}
\end{figure*}

\begin{figure}
\resizebox{\hsize}{!}{\includegraphics{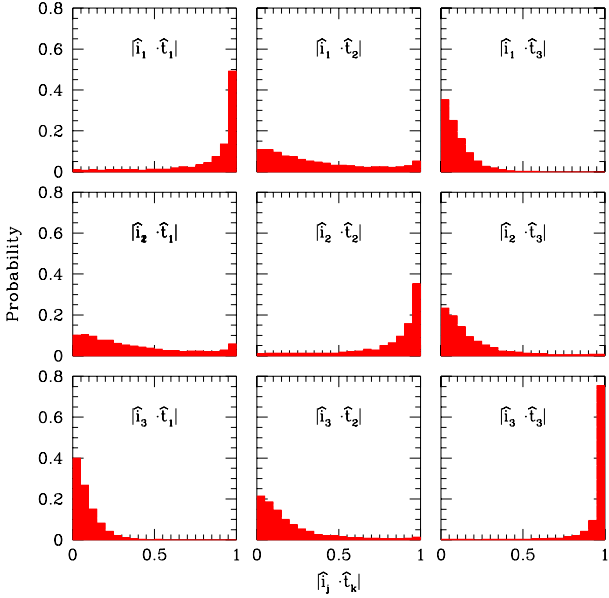}}
\caption{Alignments of the tidal torque and inertia eigentsystems, quantified by the distributions of the cosines of the angles between the principal axes of the tidal shear and inertia tensors. {\em The figure was kindly provided by C. Porciani and appears in \citet{2002MNRAS.332..325P}. Reproduced with permission from Blackwell Publishing.}}
\label{fig_alignment_it}
\end{figure}

Two popular ways of defining spin correlation functions of the angular momentum directions $\hat{L}\equiv\vecl/L$ are\begin{eqnarray}
\eta(r) & = & \langle\hat{L}(\vecx)\hat{L}(\vecx+\vecr)\rangle,\\
\eta_2(r) & = & \langle\left|\hat{L}(\vecx)\hat{L}(\vecx+\vecr)\right|\rangle - \frac{1}{3},
\end{eqnarray}
which have been measured from the $n$-body simulation by \citet{2002MNRAS.332..325P}, who find a positive spin correlation function at early times out to separations of a few comoving Mpc, but this correlation decreases towards $z=0$, due to nonlinear effects, as shown in Fig.~\ref{fig_ll_correlation}. In their second paper on the topic, \citet{2002MNRAS.332..339P} focus on misalignments between the tidal shear and inertia tensors, and investigate the extend to which the common assumption of uncorrelatedness is fulfilled. To this end, they determine the eigenvectors $\Psi_\alpha$ and $I_\alpha$ of the two tensors and measure the distribution of the cosines between two pairs of vectors. They find strong alignment between the minor axes of the two tensors, as illustrated in Fig.~\ref{fig_alignment_it}, which gives the distributions of the angles between $\Psi_\alpha$ and $I_\beta$. The correlation parameter $c_\mu$ for the directional cosines $\mu$ is found to be $c_\mu\simeq0.61$, again indicating strong correlation between shear and inertia. This correlation would affect the decomposition eqn.~(\ref{eqn_decomposition}).

A test of the models for nonlinear torquing and the resulting spin correlation functions from numerical data is provided by \citet{2007arXiv0707.1690L}. Their starting point is the observation by \citet{2002astro.ph..5512H}, that the correlation $\eta_2(r)\propto\xi^2(r)$ is only valid for Gaussian fields, and introduce a correction $\propto\xi(r)$ to the correlation function, which implies that the spin correlation is more long-ranged if nonlinear modifications to the dynamics are present. In fact, their data shows weak correlations out to $20~\mathrm{Mpc}/h$, compared to $10~\mathrm{Mpc}/h$ for linear tidal torquing, with nonlinear effects increasing with decreasing redshift, and is well fitted by the model with the additional correction term.

% --- subsection: angular momentum alignments with the cosmic web --- %
\subsection{Angular momentum alignments with the cosmic web}\label{sect_simulation_alignments}
Apart from the angular momentum correlation function, which is due to correlations in the initial conditions for angular momentum acquisition, there are alignments between angular momenta and the ambient large-scale structure caused by the large-scale tidal gravitational fields of nonlinear structures like sheets and filaments \citep[for an overview, see][]{2005MNRAS.359..272C}, which has attracted considerable interest. Due to domination of nonlinear physics, these alignments can only be investigated in $n$-body simulations. Primary topics of the analyses are, apart from angular momentum alignments, the alignments of the halo axes with the large-scale structure, and the orientation of galactic disks.

\citet{2004ApJ...613L..41N} focus on the orientation of the angular momentum axis of disk galaxies (i.e. the angular momentum of the dark matter haloes of disk galaxies) and confirm the expectation that the haloe revolves around the intermediate axis. Extending their analysis they investigate the orientations of the galactic disk and of the halo relative to the surrounding large-scale structure, they find that the orientation of rotation axis lies in a plane traced by the protogalactic material, which provides an explanation of the high inclination of Milky Way relative to the supergalactic plane.

The paper by \citet{2005ApJ...627..647B} addresses the alignment of shapes and anngular momenta inside a halo as well as relative to the ambient density field. Their results on the first point include the an alignment of the angular momentum vector with the intermediate axis of the halo ellipsoid, which shows a slight dependence on radius. The external alignment of the halo's axes and of the angular momentum was quantified by correlations between the principal axis system and the separation vector between neighbouring haloes, for which evidence was found, correlations between the principal axis systems of two haloes, which suggests correlations between the major axes only, correlations between the angular momenta and the separation vector between two haloes, with no conclusive evidence and finally the angular momentum correlation function, where no deviations from isotropy were measured. \citet{2007arXiv0709.1106L} choose to quantify shape correlations by ellipticity correlation function and ellipticity-direction correlations, and carry out measurements of those correlation functions, finding strong support on tidal torquing, and providing polynomials in the density correlation function as fitting formul{\ae}.

\citet{2006ApJ...652L..75P} and \citet{2007MNRAS.375..184B} investigate the alignment of dark matter haloes with voids in the large-scale structure and provide evidence that the minor axis of dark matter haloes points preferentially to the centres of voids, while the angular momentum was not found to have a particular orientation. Consequently, the intermediate and major axes were found to point preferentially perpendicularly to the void direction. These results apply to voids larger than $10~\mathrm{Mpc}/h$ for haloes in a shell of thickness $4~\mathrm{Mpc}/h$ around the void. The directional cosines $\mu$ between the void direction vector and the principal axes of the ellipsoid can be described by a probability distribution of the shape $P(\mu)\propto p/(1+(p^2-1)\mu)^{3/2}$, where $p=1$ corresponds to a case of no alignment.

The topic of the analysis by \citet{2007ApJ...655L...5A} is the spin alignment with nonlinear features in the large-scale structure. They use a filtering scheme for segmenting the large-scale structure into sheets and filaments, using the curvature tensor $\partial_i\partial_j\delta$ of the density field, which they reconstruct from the particle distribution with a Delauney tesselation technique. The ratio of eigenvalues of $\partial_i\partial_j\delta$ provide measures of local spherical symmetry, filamentarity or planarity: Haloes in walls have spin vectors lying in the plane of the sheet they are situated in, the orientation of spins in filaments is weaker, but acquires a mass dependence.

\citet{2007MNRAS.381...41H} investigate ellipticity and spin-parameter distributions, as well as the spin-spin and spin-orbit correlation functions, studying the environment and redshift dependence of these observables. As environmental classification they use the number of positive eignvalues of the (smoothed) tidal shear tensor, which is motivated by the fact that compression of the cosmic density field takes place along directions where the tidal shear has a positive eigenvalue. For the eigenvalues $\lambda_i$, $i=1,2,3$ of the tidal shear $\partial_i\partial_j\Psi$ one can distinguish between clusters ($\lambda_{1,2,3}\geq0$), filaments ($\lambda_{2,3}\geq 0$, $\lambda_1<0$), sheets ($\lambda_3\geq0$, $\lambda_{1,2}< 0$) and voids ($\lambda_{1,2,3}<0$). Their results comprise trends of halo sphericity and triaxiality with environmental classification, as well as enviroment-dependencies of the spin parameter $\lambda$, which on average assumes higher values in filaments for high mass haloes, and significantly lower values for low mass haloes in voids. 

\begin{figure}
\resizebox{\hsize}{!}{\includegraphics{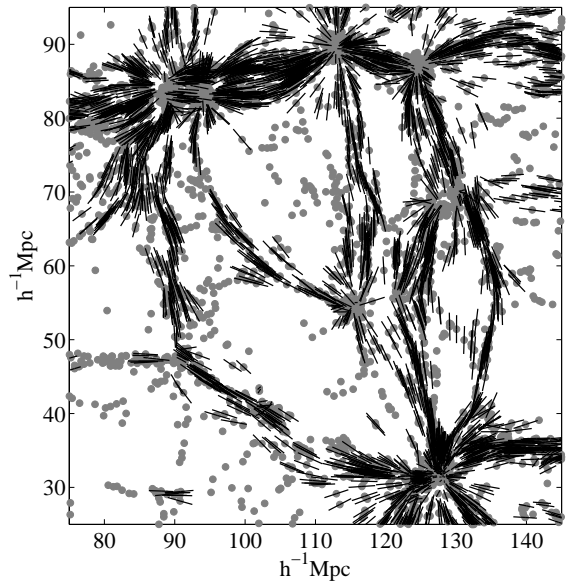}}
\caption{Slice through a simulation of the large scale density field: Grey dots indicate haloes, and the arrows denote the direction of the eigenvector corresponding to the largest eigenvalue of the tidal shear tensor inside filamentary regions, and is used for measuring shape or angular momentum alignment. {\em The figure was kindly provided by O. Hahn and appears in \citet{2007MNRAS.381...41H}. Reproduced with permission from Blackwell Publishing.}}
\label{fig_alignment_cosmic_web}
\end{figure}

Spins of low-mass haloes tend to be aligned with the filament direction, haloes similar to the nonlinear mass scale are randomly aligned whereas high-mass haloes have a weak tendency to orient their spin perpendicularly to the filament. \citet{2007MNRAS.375..489H} did not find a spin-spin correlation between neighbouring haloes, but report a substantial spin-orbit correlation, independent of mass-scale or environment: Pairs of haloes tend to align their spins parallely to the orbital angular momentum. Fig.~\ref{fig_alignment_cosmic_web} shows the tidal fields and their alignment relative to the large-scale density field in an $n$-body simulation carried out by \citet{2007MNRAS.381...41H}, indicating the important influence of nonlinear dynamics on the tidal shear.

\citet{2008MNRAS.385..867C} revisit the spin alignment with voids. They confirm the validity of the fitting formula proposed by \citet{2007MNRAS.375..184B} for the angle distribution between the ellipsoid's principal axes and the void direction ruling out random alignment with high significance, and in addition, find a preferential alignment between the angular momentum and the void direction, which was not present in earlier studies. They attribute this result to the criteria by which particles were asssigned to haloes, which impacts on the determination of the inertia tensor and of the angular momentum.

% --- angular momentum buildup by mergers --- %
\subsection{Spin-up from merging and satellite accretion}\label{sect_simulation_accretion}
Apart from tidal torquing a galaxy may acquire angular momentum by accretion of satellites. These satellites may bring in their own angular momentum or enter the merging system under a certain impact parameter, such that in the process of accretion orbital angular momentum of the satellite gets transformed into spin angular momentum of the merged system. The spin-up in satellite accretion was considered among others by \citet{2002ApJ...581..799V} and \citet{2007MNRAS.380L..58D} who find in $n$-body simulations that the mechanism of satellite accretion is sufficient to provide enough angular momentum. \citet{2007MNRAS.380L..58D} design a model for angular momentum build-up my merging, which is based on stoachastic increase in angular momentum in the sequence of merging processes of a dark matter halo, and which is able to reproduce the log-normal distribution of the spin parameter $\lambda$. Extending on the preceeding work by \citet{2004ApJ...612L..13D} who find that tidal torquing does not provide enough angular momentum for rotationally supported objects to form and conclude that accretion as a further mechanism is required. \citet{2007MNRAS.380L..58D} show that the distribution of $\lambda$ depends on the dynamical state of the halo and shows differences between relaxed haloes and haloes which underwent recent merging. The relaxation of a halo is quantified by the distance $s$ between the most bound particle and the centre of mass, which should assume small values for a relaxed, spherically symmetric object. They quantify the time evolution of $s$, $\lambda$ and halo mass. 

\citet{2004MNRAS.348..921P} take these investigations further and classify the systems according to their merging history into two categories: continuous accretion and sigificant merging. Their results suggests that the angular momentum growth exhibits a transition time at $z\simeq1.5$, where the fast initial growth levels off. In addition, they find a correlation between the mass and angular momentum evolution, suggesting that merging has a significant impact on the angular momentum build-up history. The spin parameter $\lambda$ is decreasing with time for haloes undergoing accretion, which is compensated by a small increase in merging haloes.

The numerical studies are supplemented by an analytical investigation carried out by \citet{1992MNRAS.255..729Q}, who considered angular momentum re-orientation in an accretion model. They construct spherical shells of matter, average the torque of each of those shells as a function of the density and potential, and express those two fields in terms of multipole moments, finding that the multipole series converges quickly such that only the di- and quadrupole moments contribute. They conclude that especially the dipole term, which gives rise to anticorrelated torques in neighbouring shells, and gives rise to warps as the galaxy can not maintain a common rotation axis throughout the dark matter halo.

% --- section: observations ---%
\section{observations}\label{sect_observation}
Intrinsic ellipticity correlations were measured using various statistics in a number of cosmological surveys. For instance, \citet{2000ApJ...543L.107P} measured the spin-direction correlation $\bra\left|\hat{L}(\vecx_1)\cdot\hat{r}(\vecx_2)\right|^2\ket$, $\vecr=\vecx_2-\vecx_1$, in the Tully galaxy catalogue, an all-sky survey comprising $3.5\times10^4$ galaxies out to redshifts of $z\lsim0.02$. Although the signal is weak, the measured correlation strength is consistent with theoretical estimates. The spin-spin correlation function $\bra\left|\hat{L}(\vecx_1)\cdot\hat{L}(\vecx_2)\right|^2\ket$ is measured by \citet{2001ApJ...555..106L} on a subsample of the same data set, yielding a total significance of $2.4\sigma$. They confirm that the simplified angular momentum coupling model eqn.~(\ref{eqn_ll_crittenden}) with the parameter $a\simeq0.24$ fitting the data nicely.

\citet{2002MNRAS.333..501B} set out to measure the alignment of elliptical galaxies in the SuperCOSMOS data set, which reaches out to a median redshift $z_\mathrm{med}=0.1$ and comprises $2\times10^6$ galaxies over a quarter of the sky. As a statistical quantifyer, they use the variance of the obseved ellipticity, and show that it is larger than the expected lensing signal, and well in accordance with the predicition from quadratic alignment models, in particular with \citet{2001ApJ...559..552C}.

Another high-significance measurement of the spin-shear correlation is reported by \citet{2002ApJ...567L.111L}, who constructed both the spin field as well as the tidal shear field (via the density field) from the PSCz-catalogue, a complete redshift catalogue from the IRAS-survey covering 84\% of the sky, and directly quantified the correlation between the two fields. The null-hypothesis of random alignments is rejected at 99.98\% confidence, and the alignment parameter was found to be $a\simeq0.17\pm0.04$, slightly smaller than the value suggested by numerical simulations, although the authors emphasis that their data treatment is likely to systematically underestimate $a$.

\citet{2006ApJ...640L.111T} perform a measurement of the orientation of the rotation axes of galaxies situated on the shells of large voids, using the 2dFGRS and SDSS data sets. They measure the distribution of the angle between spin axis and the density gradient (i.e. a related quantity to the void centre direction) and can reject the null hypothesis of randomly aligned galaxies at 99.7\% confidence. Comparing their distribution to the prediction derived from eqn.~(\ref{eqn_ll_crittenden}), the data requires much larger values for $a$, $a\simeq0.7$, but provides otherwise an excellent fit.

\citet{2007ApJ...671.1248L} combine a reconstruction of the density respective tidal shear field from the 2MASS reshift survey with the spin field derived from the Tully galaxy catalogue. The measured correlation strength exceeds $6\sigma$ significance, and allow the investigation of the morphology and environment dependence. \citet{2007ApJ...671.1248L} find a weak increasing trend of the correlation with increasing morphological type and measure larger correlations in high-density environments. Additionally, they confirm the predicitions of tidal torquing to be well describing the data, in particular the probability desity distribution of the cosine between the spin and the intermediate axis of the tidal shear tensor. Their mean correlation parameter $a\simeq0.084\pm0.014$ is much smaller than the value obtained in numerical simulations.

Aiming at distinguishing different alignment scenarios for elliptical and spiral galaxies, \citet{2007ApJ...670L...1L} carry out an investigation of alignments of red and blue galaxies in SDSS, selecting galaxies out to redshifts of $z=0.4$. For the description of intrinsic shape correlation the functional form $\eta(r)\propto a\xi^2(r)+\epsilon\xi(r)$ is used, where the case $a\ll\epsilon$ would correspond to the usual quadratic alignment model for spiral galaxies with a small correction due to non-Gaussianity, whereas $a\gg\epsilon$ would be applied to elliptical galaxies, where a linear alignment model applies. The measurement in fact supports this ideas as it yields $a=0.20\pm0.04$ and $\epsilon\simeq0$ for blue galaxies and $a\simeq0$ and $\epsilon=(2.5\pm0.5)\times10^{-3}$ for red galaxies.

% --- section: lensing ---%
\section{implications for gravitational lensing}\label{sect_lensing}
An important application of angular momentum correlation models is the problem of intrinsic alignments in weak gravitational lensing. After a short recapitulation of gravitational lensing (Sect.~\ref{sect_lensing_lensing}) and especially weak cosmic shear (Sect.~\ref{sect_lensing_wl}), I explain the contamination of lensing data by intrinsic aligments in the measurement of weak shear spectra (Sect.~\ref{sect_lensing_ia}) and review common models linking ellipticity correlations to angular momentum correlations (Sect.~\ref{sect_lensing_ellipticity}). Then, I discuss measurements of the intrinsic alignment terms, both on simulations as well as on actual data  (Sect.~\ref{sect_lensing_measurement}) and give an overview over methods proposed to suppressing intrinsic alignments (Sect.~\ref{sect_lensing_suppression}).

% --- subsection: gravitational lensing --- %
\subsection{Gravitational lensing}\label{sect_lensing_lensing}
Gravitational lensing \citep[for a review, see][]{2001PhR...340..291B} refers to the deflection of light reaching the observer from a distanct galaxy by intervening gravitational potentials. The position $\bbeta$ at which a galaxy is observed is related to the angle $\btheta$ under which the galaxy would be observed without gravitational lensing by the lens equation,
\begin{equation}
\bbeta = \btheta - \balpha(\btheta),
\end{equation}
with the deflection field $\balpha(\btheta)$, which is given as the angular gradient $\balpha=\nabla\psi$ of the deflection potential $\psi$. $\psi$ follows from the surface mass density $\Sigma(\btheta)$ of the lens as a solution to the 2d Poisson equation,
\begin{equation}
\Delta\psi(\btheta) = 2\kappa(\btheta) \rightarrow 
\psi(\btheta) = \int\dd^2\theta^\prime\: \kappa(\btheta^\prime) \ln\left|\btheta-\btheta^\prime\right|,
\end{equation}
after normalisation with the surface critical density $\Sigma_\mathrm{crit}=c^2/(4\pi G)D_s/(D_d D_{ds})$ such that the convergence $\kappa$ is related to the surface mass density of the lens via $\kappa(\btheta)=\Sigma(\btheta)/\Sigma_\mathrm{crit}$. The angular diameter distance to the lensed galaxy is denoted by $D_s$, and likewise $D_d$ is the angular diameter distance to the lens and $D_{ds}$ the distance between lens and background galaxy. Since $\btheta$ can principally never be known, one rather observes linear variations of the deflection field $\balpha$ across the image of the galaxy leading to shape changes in the galaxy image. This information is contained in the Jacobian $\mathcal{A}$,
\begin{equation}
\mathcal{A}\equiv\frac{\partial\bbeta}{\partial\btheta},\quad
\mathcal{A}_{ij} = \delta_{ij} - \frac{\partial^2\psi}{\partial\theta_i\partial\theta_j}
\quad\rightarrow\quad
\Delta\mathcal{A}_{ij} \equiv \frac{\partial^2\psi}{\partial\theta_i\partial\theta_j}
\end{equation}
which describes the shape distorsions of an image to first order. The difference $\Delta\mathcal{A}$ between the Jacobian and the unit-mapping can be decomposed with the Pauli-matrices $\sigma_\beta$,
\begin{equation}
\Delta\mathcal{A}
= \sum_{\beta=0}^3\: a_\beta\sigma_\beta 
= \kappa\sigma_0 + \gamma_+\sigma_1 - \ci\rho\sigma_2 + \gamma_\times\sigma_3,
\end{equation}
because the $\sigma_\beta$ constitute a complete basis of the vector space of the real $2\times 2$-matrices \citep{2005mmp..book.....A}. The coefficients are refered to as the convergence $\kappa$, which describes the isotropic image size change, the two components of shear $\gamma_+$ and $\gamma_\times$, and the image rotation $\rho$, which can only be excited by multiple lensing events along the line of sight or if corrections due to the perturbed photon geodesic become appreciable. Due to the property $\sigma_\alpha\sigma_\beta = \mathrm{id}(2)\delta_{\alpha\beta} + \ci\epsilon_{\alpha\beta\gamma}\sigma_\gamma$ of the Pauli-matrices, the coefficients can be recovered using $a_\beta=\trace(\Delta\mathcal{A}\:\sigma_\beta)/2$. Due to their common transformation under coordinate system rotations, the two shear components can be combined to the complex ellipticity $\gamma = \gamma_+ + \ci\gamma_\times$, which forms a spin-2 field and transform according to $\gamma\rightarrow\gamma\exp(2\ci\varphi)$ in rotations about an angle $\varphi$.

The shape of a galaxy is measured from the second moments of the brightness distribution $Q_{ij}$, which in turn defines the complex ellipticity $\epsilon$,
\begin{equation}
\epsilon = \epsilon_+ +\ci\epsilon_\times,\quad\mathrm{with}\quad
\epsilon_+=\frac{Q_{11}-Q_{22}}{Q}\quad\mathrm{and}\quad\epsilon_\times=\frac{2Q_{12}}{Q},
\end{equation}
where I abbreviated $Q=Q_{11}+Q_{22}+2({Q_{11}Q_{22}-Q_{12}^2})^{1/2}$. The complex ellipticity forms a spin-2 tensor field due to the transformation property $\epsilon\rightarrow\epsilon\exp(2\ci\varphi)$ under the rotation of the coordinate system by an angle $\varphi $ and maps onto itself if $\varphi =\pi$. Under the action of gravitational lensing, the intrinsic ellipticity $\epsilon_s$ of a galaxy is distorted to the observed ellipticity $\epsilon$ given by \citep{1997A&A...318..687S}
\begin{equation}
\epsilon = \frac{\epsilon_s + g}{1 + g^*\epsilon_s},
\end{equation}
with the complex reduced shear $g\equiv\gamma/(1-\kappa)$. In weak lensing applications the relation simplifies to $g\simeq\gamma$ because $\left|\kappa\right|\ll 1$.

% --- subsection: weak shear --- % 
\subsection{Weak cosmic shear}\label{sect_lensing_wl}
Weak gravitational lensing directly probes the fluctuations of the cosmic density field by imprinting correlated changes to the shapes of neighbouring galaxies, because the light rays reaching us from pairs of galaxies have to transverse the same part of the large-scale structure and experience correlated tidal fields. In this context, the term weak refers to the fact that the distorsions imprinted onto the galaxy shapes are small, $\kappa\ll 1$, $\left|\gamma\right|\ll 1$, and the signal is obtained as a statistical average over many galaxies. Although the primary observable is the shearing of the galaxy image, one often works in terms of the lensing convergence $\kappa$ as it has the same statistical properties as the shear and is easier to work with.

The convergence $\kappa$ is given by a line of sight integration:
\begin{equation}
\kappa = \frac{3H_0^2\Omega_m}{2c^2}\int_0^{\chi_H}\dd\chi\: G(\chi)\chi\:\frac{D_+}{a}\delta,
\label{eqn_wl_convergence}
\end{equation}
replacing the actual photon geodesic by a straight line (Born-approximation), with the lensing-efficiency weighted distribution of galaxies $G(\chi)$,
\begin{equation}
G(\chi) = \int_\chi^{\chi_H}\dd\chi^\prime\: p(\chi^\prime) \frac{\chi^\prime-\chi}{\chi^\prime}.
\end{equation}
The redshift distribution of the lensing galaxies is denoted by $p(\chi)\dd\chi$, after changing the distance measure from redshift to comoving distance. The angular power spectrum $C_\kappa(\ell)$ of the lensing observable is related to the power spectrum $P(k)$ of the density field via Limber-projection \citep{1954ApJ...119..655L},
\begin{equation}
C_\kappa(\ell) = \frac{9H_0^4\Omega_m^2}{4c^4}\int_0^{\chi_H}\dd\chi\: G^2(\chi)
\frac{D_+^2(a)}{a^2} P(k=\ell/\chi),
\end{equation}
where the integration range formally extends to the comoving horizon distance $\chi_H$. Technically, one can tremendously improve the sensitivity of weak lensing by dividing the galaxy sample into redshift bins and carry out weak lensing tomography, where the lensing spectrum is measured separately from each redshift bin containing the respective information on structure growth and the geometrical factors, and by combining those measurements. The weak lensing convergence $\kappa_i$ in a tomographic measurement reads
\begin{equation}
\kappa_i = \int_0^{\chi_H}\dd\chi\:G_i(\chi)\chi \frac{D_+}{a}\delta,
\end{equation}
with the altered lensing efficiency function $G_i(\chi)$
\begin{equation}
G_i(\chi) = \int_\chi^{\chi_H}\dd\chi^\prime\:p_i(\chi^\prime)\frac{\chi^\prime-\chi}{\chi^\prime},
\end{equation}
constructed for the galaxy distance distribution $p_i(\chi)\dd\chi$ of the redshift bin $i$. In this way, on can access cosmological information that would be otherwise diluted by the line of sight integration eqn.~(\ref{eqn_wl_convergence}) with a slowly varying weighting function. A further development of this idea proposed by \citet{2006MNRAS.373..105H} is refered to as 3d-lensing, where the density field and its time evolution is reconstructed from the weak lensing shear, which is effectively a measurement of the local tidal field.

% --- subsection: intrinsic alignments --- % 
\subsection{Intrinsic alignments}\label{sect_lensing_ia}
A common assumption in weak lensing are uncorrelated intrinsic shapes, which is challenged by the fact that there are ellipticity correlations as a consequence of angular momentum correlations. In the case of uncorrelated intrinsic ellipticities, the observed angular convergence power spectrum $\tilde{C}_\kappa(\ell)$ differs from the theoretically expected spectrum $C_\kappa(\ell)$ by a Poissian noise term $\sigma_\epsilon^2/n$, which eventually dominates on small angular scales,
\begin{equation}
\tilde{C}_\kappa(\ell) = C_\kappa(\ell) + \frac{\sigma_\epsilon^2}{n}.
\end{equation}
The dispersion of the Gaussian ellipticity distribution is usually quoted to be $\sigma_\epsilon=0.4$, and $n$ denotes the number of galaxies per steradian. Typical number densities of galaxies range between 20 and 100 objects per squared arcminute, depending on the survey setup. If the galaxy shapes are intrinsically correlated, the above relation becomes
\begin{equation}
\tilde{C}_\kappa(\ell) = C_\kappa(\ell) + C_\epsilon(\ell).
\end{equation}
with the intrinsic ellipticity correlation $C_\epsilon(\ell)$ if the intrinsic shapes are uncorrelated with the lensing shear. This is actually not the case, because the same tidal shear field gives rise to both the intrinsic ellipticity alignment as well as the lensing signal, i.e. there can be an additional cross correlation term $C_{\kappa\epsilon}(\ell)$ \citep{2004PhRvD..70f3526H},
\begin{equation}
\tilde{C}_\kappa(\ell) = C_\kappa(\ell) + C_\epsilon(\ell) + C_{\kappa\epsilon}(\ell),
\end{equation}
especially in linear alignment models. To be exact, $C_{\kappa\epsilon}(\ell)$ describes the correlation between the lensing induced apparent shape change of a background galaxy and the physical alignment of a foreground galaxy in the tidal field of the gravitational lens. The physical shape of a background galaxy can safely be assumed to be uncorrelated with the tidal field of the lens as they are typically separated my many correlation lengths.

The quantification of intrinsic ellipticity correlations in SDSS and the implications for weak lensing is the topic of the paper by \citet{2006MNRAS.367..611M}: Using a spectroscopic sample of $2.6\times10^5$ galaxies at low redshift ($z\lsim0.12$), they carry out measuremets of the intrinsic ellipticity correlation $C_\epsilon(\ell)$ as well as of the cross correlation $C_{\kappa\epsilon}(\ell)$ between ellipticity and density. While the first correlation was too week to be detected, the latter correlation, present in linear alignment models, yielded a clear detection in bright galaxies out to scales of $60~\mathrm{Mpc}/h$, with the correct sign. Inferring consequences for weak lensing from their results suggests that the contamination of deep lensing surveys with intrinsic alignments could be as large as 10\%, and that values of $\sigma_8$ derived from weak lensing might be overestimated by as much as 20\%, if uncorrected. Additionally, they provide fits to the alignment model proposed by \citet{2000MNRAS.319..649H} for carrying out that correction.

Extending this analysis, \citet{2007MNRAS.381.1197H} investigate correlations between ellipticity and the density field, using $3.6\times10^5$ LRGs from SDSS and 8000 LRGs from 2SLAQ, resulting in a $>3\sigma$ detection of the alignment effect on large scales up to $60~\mathrm{Mpc}/h$. The high detection significance allows the determination of fitting functions for the scaling of alignments in relation to luminosity, transverse separation and redshift. Applying their findings to the estimation of cosmological parameters from weak lensing surveys they conclude that the contamination can decrease the lensing spectrum by typically 6.5\%, which induces as bias in the estimation of $\sigma_8$ of $\Delta\sigma_8=-0.02$. 

\citet{2007NJPh....9..444B} show that the intrinsic alignment contributions to weak lensing spectra cause severe errors in the estimation of cosmological parameters, in particular the dark energy equation of state parameter $w$, which can be biased by as much as 50\% if the alignments remain uncorrected. \citet{2007NJPh....9..444B} emphasise the importance of accurate photometry in tomographic weak shear measurements, with respect to the total number of redshifit bins as well as the photometric redshift accuracy.

% --- subsection: ellipticity correlations --- %
\subsection{Ellipticity correlations}\label{sect_lensing_ellipticity}
A consequence of angular momentum correlations between neighbouring spiral galaxies are the correlations between their apparent ellipticities, because their galactic disks are viewed under similar angles of inclination. In doing that, one assums that the galactic disks form perpendicular to the angular momentum axis of the host halo. For disks with zero thickness the relation between the scalar ellipticity $\epsilon$ and the line of sight component $L_z$ of the angular momentum is given by
\begin{equation}
\epsilon = \frac{1-\hat{L}_z^2}{1+\hat{L}_z^2}.
\end{equation}
Being a purely geometrical effect, the ellipticity depends only on the direction $\hat{L}=\vecl/L$ and not on the absolute value of the angular momentum. The two components of the complex ellipticity $\epsilon = \epsilon_+ + \ci\epsilon_\times$ can be derived from the angular momentum direction via
\begin{equation}
\epsilon_+ = \frac{\hat{L}^2_x - \hat{L}^2_y}{1+\hat{L}^2_z}
\quad\mathrm{and}\quad
\epsilon_\times = 2\frac{\hat{L}_x\hat{L}_y}{1+\hat{L}^2_z},
\end{equation} 
such that $\epsilon^2=\epsilon_+^2+\epsilon_\times^2$. This alignment model is commonly refered to as the quadratic alignment model, due to the dependence on $\hat{L}^2$. Contrarily, the alignment model applying to elliptical galaxies is usually parameterised to be linear in $L$. 

A finite thickness of the galactic disc can be incorporated by a constant of proportionality $\alpha$ in the range $0<\alpha<1$, 
\begin{equation}
\epsilon = \alpha\frac{1-\hat{L}_z^2}{1+\hat{L}_z^2}.
\end{equation}
which weakens the dependence of $\epsilon$ on $\hat{L}_z$. \citet{2001ApJ...559..552C} measure $\alpha$ to have a value of $\alpha\simeq 0.85$ for spirals, and $\alpha\simeq 0.5$ for spheroids and ellipticals. For typical mixture between spiral, lenticular and elliptical galaxies from the APM-survey, they obtain a mean value of $\alpha\simeq 0.73$.

\begin{figure*}
\begin{tabular}{cc}
\resizebox{9cm}{!}{\includegraphics{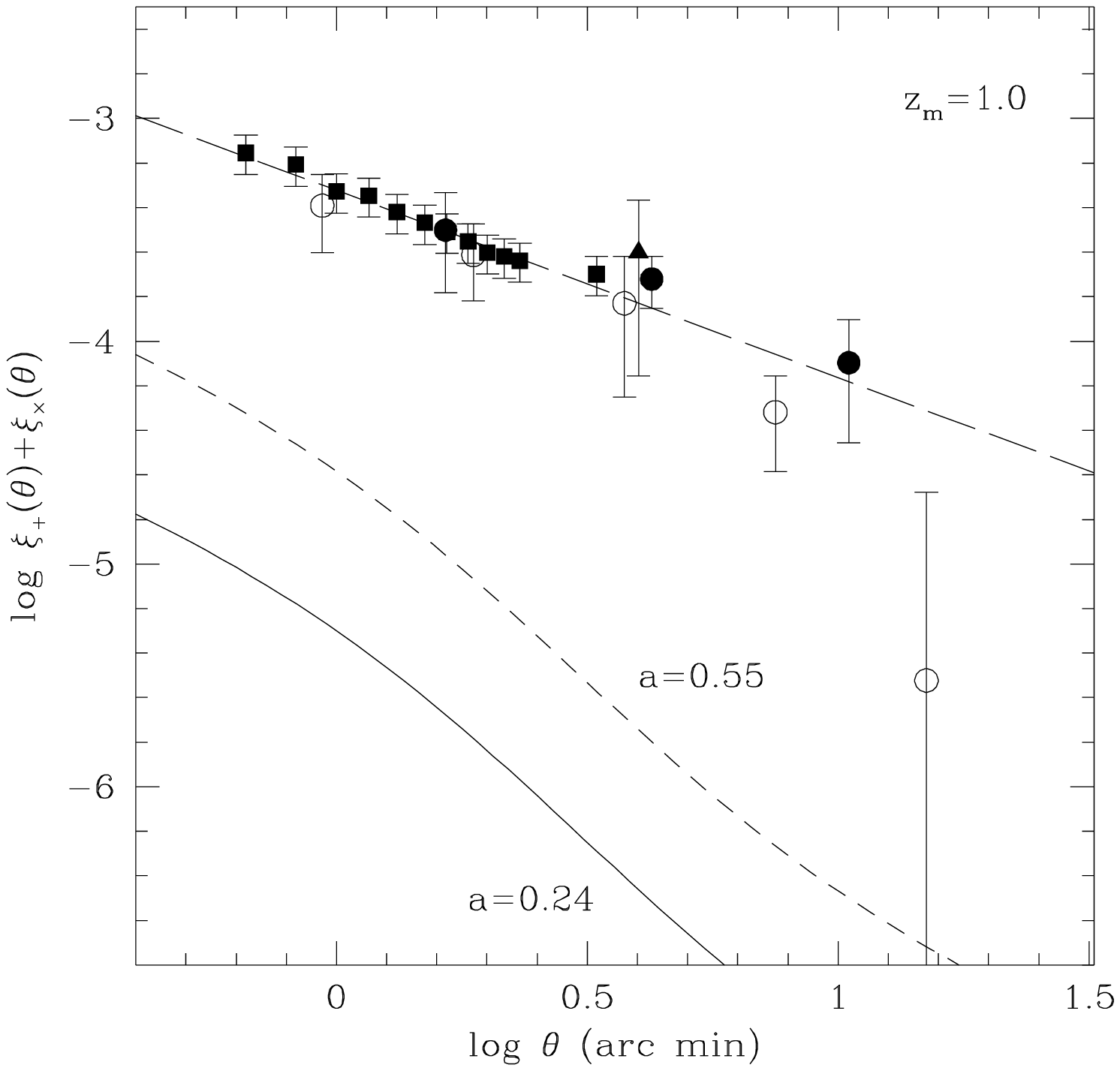}} &
\resizebox{9cm}{!}{\includegraphics{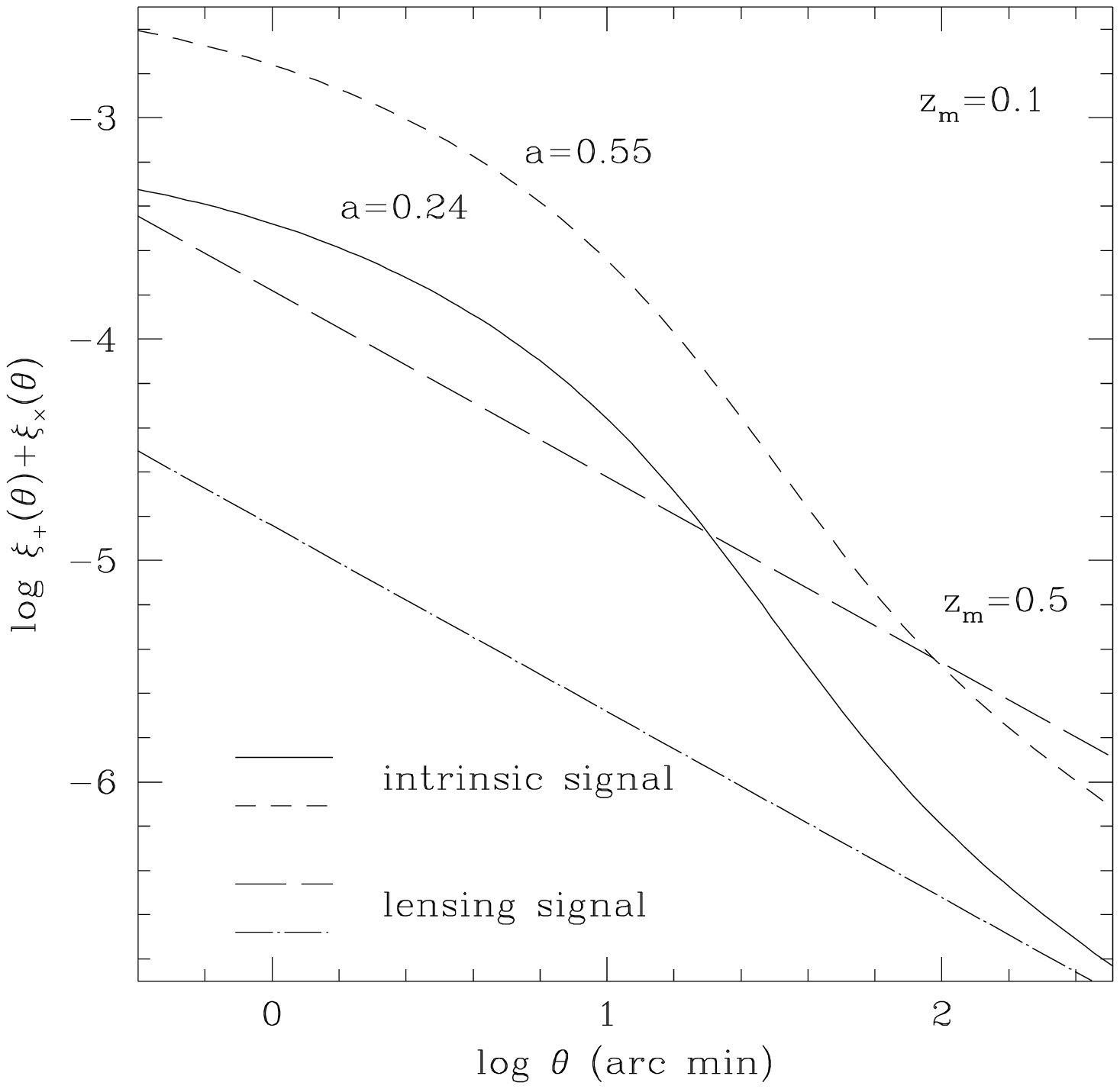}}
\end{tabular}
\caption{Comparison of the weak lensing shear and the intrinsic alignment signal for deep surveys ($z_\mathrm{med}=1$, left panel), and for shallow surveys ($z_\mathrm{med}=0.1$, right panel). The intrinsic alignment signal is predicted using a quadratic alignment model with a spin correlation parameter of $a=0.24$ (solid line) and $a=0.55$ (dashed line). The long-dashed line indicates the intrinsic lensing signal, with superimposed data point from lensing surveys. {\em The figure was kindly provided by R. Crittenden and appears in \citet{2001ApJ...559..552C}. Reproduced with permission from the American Astronomical Society.}}
\label{fig_intrinsic}
\end{figure*}

After linking the ellipticity correlations to angular momentum correlations, \citet{2001ApJ...559..552C} continue by using the decomposition eqn.~(\ref{eqn_decomposition}) along with multivariate Gaussians for the conditional probability density $P(\vecl|\bmath{\Psi})$ and the quadratic model eqn.~(\ref{eqn_ll_crittenden}) for the relation between angular momentum and shear for deriving the 2-point correlation function of the ellipticity field. They confirm the important result that the ellipticity correlation function is proportional to the squared correlation function $\xi(r)$ of the density field. In the next step they use a Limber-projection with a weighting proportional to the galaxy density including galaxy clustering in order to obtain the angular correlation function of the intrinsic ellipticity, which then can be compared to the angular correlation function of the weak lensing shear. Intrinsic alignments were found to be a serious contaminant only for shallow surveys, as illustrated in Fig.~\ref{fig_intrinsic}: A survey with median redshift of 0.1 would be dominated by the intrinsic alignment signal, whereas the intrinsic ellipticity alignments are smaller by an order of magnitude for a deep survey with a median survey redshift of unity. \citet{2000ApJ...532L...5L} take this result to extremes and show that the density field could be reconstructed from the ellipticity field due to the angular momentum distribution instead of using the tidal field measured by weak lensing, by deriving a formula for the spin-density correlation.

In a related work, \citet{2002MNRAS.332..788M} investigate angular momenutum-induced ellipticity correlations, using the same quadratic model as \citet{2001ApJ...559..552C}, derive the angular power spectra and compare the signal amplitude to the lensing power spectra. Technically, they treat the tidal field as being continuous and the inertia and shear as uncorrelated. Comparing the intrinsic alignment to the cosmological lensing signal, they confirm the dominance of the intrinsic alignments over the lensing for surveys with low source redshifts of $z\simeq0.1$, equality of both effects for redshifts of $z\simeq0.3$ and show that a deep survey with $z\simeq1$ has a negligible contamination of the lensing signal by more than two orders of magnitude.

In what concerns galaxy-galaxy lensing, where the foreground galaxy density is correlated against the background ellipticity field at small angular separations, the impact of intrinsic alignments is investigated in detail by \citet{2002astro.ph..5512H}. Their starting point is the acute observation that tidal models, where the alignment is proportional to the squared tidal shear, can not cause a non-vanishing ellipticity-galaxy density cross-correlation $\bra\gamma\epsilon\ket$: The correlator could be expanded into a 3-point function, which would vanish, provided that the density field is Gaussian, the bias is unity and there would be no source clustering. They continue by quantifying the magnitude of the correlation $\bra\gamma\epsilon\ket$ for non-Gaussianities which arise in the course of structure formation. While the result is very interesting for cosmological weak lensing and for testing intrinsic alignment models in themselves, they bridge the gap to galaxy-galaxy lensing by showing that there are influences by intrinsic alignments because the separation into fore- and background objects is imperfect. An investigation of the transition between galaxy-galaxy lensing and cosmic weak shear is provided by \citet{2007ApJ...655L...1B}.

% --- subsection: quantification of the intrinsic alignment signal --- %
\subsection{Quantification of the intrinsic alignment signal}\label{sect_lensing_measurement}
One of the first studies for quantifying intrinsic alignment in comparison to the cosmological weak lensing signal was carried out by \citet{2000MNRAS.319..649H}, who measured the ellipticity spectrum $C_{\epsilon}(\ell)$ from haloes in an $n$-body simulation, assuming that galactic disks are aligned with the angular momentum of the host halo. Quite independent of cosmology, \citet{2000MNRAS.319..649H} found that the weak shear signal $C_\kappa(\ell)$ dominates over $C_{\epsilon}(\ell)$ by about an order of magnitude, for a deep survey, but show that the reverse applies to shallow surveys, \spirou{for angular momentum aligned spiral galaxies. For the case of elliptical galaxies, which are assumed to have the same axis system as the host halo, they report the measurement of an ellipticity correlation function with an amplitude twice as large as for spiral galaxies, but show that the lensing signal of a deep survey dominates over this intrinsic ellipticity correlation}.

Using a very different model, \citet{2001MNRAS.320L...7C} provide a theoretical discussion of intrinsic alignments in comparison to the weak lening signal, by assuming that the galaxies are distorted physically by the tidal shear field, which causes the ellipticity to be proportional to the tidal shear and not the square. They recover the result that deep surveys are almost free from intrinsic alignment contamination, as the spectra differ by almost two orders of magnitude, and that intrinsic alignments become appreciable in shallow surveys. Comparing their alignment model with alignments due to angular momentum correlations they emphasise that angular-momentum induced ellipticity alignments, being proportional to the square of the tidal shear, are short-ranged compared to alignments directly caused by ellipticity deformations due to the tidal shear. 

\citet{2000ApJ...545..561C} measure the correlation of ellipticities of dark matter haloes in an $n$-body simulation, assuming that the halo shapes serve as proxies for the galaxy shapes, which are measured from the second moments of the projected mass distribution. They report a positive correlation out to distances of $20~\mathrm{Mpc}/h$, and find that the signal is not strongly affected by varying the halo-finding technique (in particular the friends-of-friends linking length), nor by the numerical resolution. In comparing the intrinsic ellipticity correlation to the weak lensing signal, they show that the ellipticity correlation is a small contamination for deep surveys, but mimicks the functional behaviour of the lensing signal. Additionally, they confirm that the first reports on measurements of cosmic shear actually pick up the cosmological weak lensing signal, and that the intrinsic alignment contamination amounts to roughtly 10\% in these surveys \citep{2000MNRAS.318..625B, 2000Natur.405..143W, 2000astro.ph..3338K, 2000A&A...358...30V}.

Again working with $n$-body data, \citet{2002MNRAS.335L..89J} developed intrinsic alignment studies for weak lensing applications: The observed ellipticity was assumed to be determined by the halo ellipticity, which allows the direct measurement of the ellipticity correlation function as well as of the angular ellipticity correlation function. Placing the haloes at unit redshift and comparing with the lensing convergence correlation function, \citet{2002MNRAS.335L..89J} confirmed that intrinsic alignments may be a serious weak lensing contaminant. It should be emphasised, however, that in these early studies an alignment model applicable to elliptical galaxies was used, which was later shown to generate a significantly higher correlation signal compared to a model based on angular momentum alignments, which is a better choice for spiral galaxies, because they make up the majority of field galaxies. In particular, an alignment model based purely on halo-ellipticity would be already ruled out by weak lensing observations, as it overpredicts the signal.

The work by \citet{2004PhRvD..70f3526H} compares the magnitude of intrinsic ellipticity correlations and ellipticity-shear cross correlations in a simplified linear alignment model, and investigate the dependence on source redshift. The usage of the linear alignment model is motivated by the fact that quadratic models do predict the cross-correlation between ellipticity and shear to vanish: In the framework of Gaussian fluctuations, one would need to compute the expectation value of an odd moment of the density field, which vanishes exactly. In an analytical calculation they derive that in the linear alignment model the primary contaminant of the weak lensing signal is the ellipticity-shear cross orrelation for almost the entire multipole range, and is typically larger by an order of magnitude compared to the intrinsic ellipticity correlation, too. 

\citet{2006MNRAS.371..750H} focus on the quantification of the contamination of weak lensing measurements due to intrinsic alignments from simulations. They identify haloes in a cosmological $n$-body simulation and model the appearance galaxies in these haloes, either by the ellipticity of the halo for an elliptical galaxy or by the inclination angle of the angular momentum axis relative to the line of sight, allowing for a misalignment between angular momentum and symmetry axis of the disk. The derived ellipticity correlation function $\eta_\epsilon(r)$ can be parameterised by a model,
\begin{equation}
\eta_{\epsilon}(r) = \frac{A}{1+(r/B)^2}
\end{equation}
with two parameters $A$, $B$, which agrees well with the alignments measurements carried out by \citet{2006MNRAS.367..611M} for the spirals, but shows deviations in the elliptical galaxy set. Morphologically, the intrinsic ellipticity correlation assumes the highest values for the elliptical galaxies and much lower values for spirals. 

\begin{figure}
\resizebox{\hsize}{!}{\includegraphics{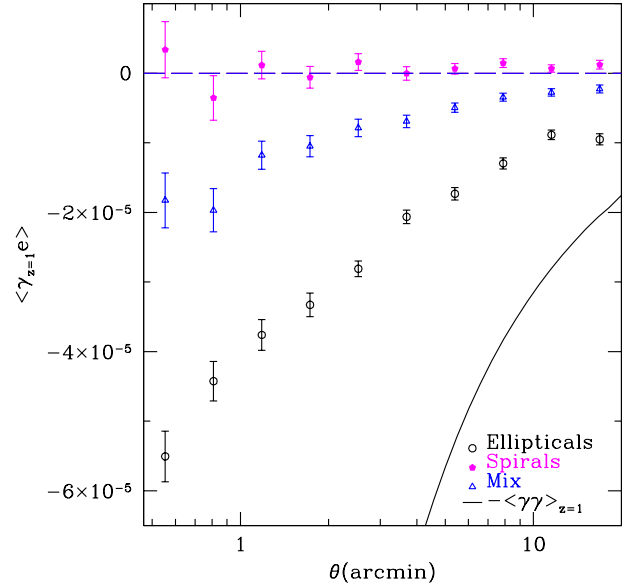}}
\caption{Lensing-intrinsic shape cross correlation functions derived from $n$-body simulations, for spiral galaxies (filled dots), a mixed population (triangles) and ellipticals (circles), in comparison to the lensing signal (solid line), which is modeled by placing the source galaxies at a redshift of unity. {\em The figure was kindly provided by C. Heymans and appears in \citet{2006MNRAS.371..750H}. Reproduced with permission from Blackwell Publishing.}}
\label{fig_lensing}
\end{figure}

At the same time, they derive the cosmological weak lensing signal from the weighted projected surface mass density of the simulated density field, and distinguish between intrinsic ellipticity correlations, which can be dealt with if accurate photometry is available by discarding close pairs, and correlations between the intrinsic ellipticity and the external gravitational shear. The latter correlation can be equally modelled by a simple function, depending on the angle of separation $\theta$,
\begin{equation}
\xi_{\epsilon\kappa}(\theta) = \frac{D_d D_{ds}}{D_s}\frac{A}{\theta + \theta_0},
\label{eqn_gi_heymans_model}
\end{equation}
with two parameters $A$ and $\theta_0$, and scaling with the lensing efficiency. As illustrated by Fig.~\ref{fig_lensing}, the correlation $\xi_{\epsilon\kappa}$ is smaller compared to the weak lensing signal by a factor of a few in a deep survey with sources at unit redshift, while ellipticals show the strongest signal, followed my a realistic mix of morphologies. On the contrary, the correlation function is consistent with zero for a data set comprising only spiral galaxies.

The impact of intrinsic alignments on higher-order statistics of the lensing signal was investigated by \citet{2008arXiv0802.3978S}. Using the same alignment model as \citet{2006MNRAS.371..750H}, they report a surprisingly strong contamination of the weak shear 3-point correlation function. Specifically, they consider the skewness of the aperture-mass statistic, where the lensing signal is weighted inside a circular aperture, and compare the three possible 3-point functions between the intrinsic alignment signal and the cosmological weak lensing signal, based on nummerical data. In shallow surveys with median redshifts of $z_\mathrm{med}=0.4$, one expects the 3-point correlation of the intrinsic alignment to be larger than the weak lensing by almost an order of magnitude, whereas in medium-deep surveys with $z_\mathrm{med}=0.7$ the contamination amounts to $\sim15\%$. Consequently, \citet{2008arXiv0802.3978S} emphasise the usability of 3-point statistics to test intrinsic alignment models. 

\spirou{\citet{2008A&A...479....9F} estimate the magnitude of the intrinsic alignment contamination on the CFHTLS weak lensing survey, by using the model for $\xi_{\kappa\epsilon}(\theta)$ \citep[see eqn.~\ref{eqn_gi_heymans_model}, introduced in][]{2006MNRAS.371..750H}. Setting the angular scale to $\theta_0=1$~arcmin, a combined fit to the cosmological and intrinsic alignment signal yields the value $A=2.2^{+3.8}_{-4.6}$ for the amplitude $A$, i.e. compatible with zero. In the range of angular scales considered, the weak lensing signal dominates over the intrinsic alignment signal and the contamination is weak, limited to be less than 32\% and larger than -13\%.}

% --- subsection: suppression of intrinsic ellipticity alignments --- %
\subsection{Suppression of intrinsic ellipticity alignments}\label{sect_lensing_suppression}
The methods for suppressing intrinsic alignment contamination in the cosmological weak shear signal fall roughly into five categories: With accurate models for the ellipticity correlations, the intrinsic alignment signal can be modelled such that one obtains a prediction for $C_\kappa(\ell)+C_{\kappa\epsilon}(\ell)+C_\epsilon(\ell)$ for a given cosmology, and it becomes possible to fit the entire observational data set. Sadly, the models for the computation of $C_{\epsilon}(\ell)$ and of the cross term $C_{\kappa\epsilon}(\ell)$ are not yet reliable enough. Secondly, one can take advantage of particular statistical properties of the intrinsic alignment signal, e.g. the excitation of vortical patterns in the ellipticity field and statistically separate the weak shear signal from the intrinsic alignment signal. A third method is the truncation of data, i.e. the rejection of galaxy pairs with small line of sight separation, at the expense of reducing statistics, especially at small separations. A fourth way of dealing with the problem is the introduction of new lensing weighting functions, which null out the intrinsic signal. At last, it is possible to treat the intrinsic alignment signal as a systematic, leave it in the data and quantify the bias in the cosmological parameter estimation which is due to the presence of intrinsic alignment. \spirou{Most of the proposed methods focus on eliminating the contribution $C_\epsilon(\ell)$, because it was only realised by \citet{2004PhRvD..70f3526H}, that the $C_{\kappa\epsilon}(\ell)$ might play an important role. Generally, the methods only assume that the intrinsic alignment contamination is local, and can be expected to treat $C_\epsilon(\ell)$ better than $C_{\kappa\epsilon}(\ell)$, due to the shorter correlation length.}

For suppressing intrinsic alignments, \citet{2002ApJ...568...20C} put forward a decomposition of the excitations in the ellipcitity field into its gradient and a curl components, which is motivated by the fact that gravitational lensing can only excite the gradient component and leaves the curl component untouched, as long as one assumes the validity of the Born-approximation and the uncorrelatedness of multiple lensing events along the line of sight. Furthermore, the lensing signal is linear in the tidal shear, whereas the intrinsic alignments measure the square of the tidal shear. \citet{2002ApJ...568...20C} show that the two contributions to the ellipticity correlations can be statistically disentangled, and emphasise the importance of measuring the curl components as a tool for controlling systematics in the observation. \citet{2002MNRAS.332..788M} investigate the ratio of the gradient and curl power spectra, and find a typical enhancement of the gradient power spectrum over the spectrum of the vortical excitations of roughly 3.5 on small angular scales. Hence, the contamination of the weak lensing spectrum is significantly larger than one would na{\"i}vely expect from a determination of the amplitude of vortical modes in the ellipticity field due to intrinsic alignment.

\citet{2002A&A...396..411K} propose a weighting scheme for directly downweighting close pairs of galaxies, where the shape correlations are likely due to intrinsic alignments and not of cosmological origin. This is justified by the fact that tidal torquing models predict only short ranged correlations, due to their dependence on the squared tidal field. In the computation of the shear correlation function they propose to use a redshift weighting $Z(z_i,z_j)$,
\begin{equation}
Z(z_i,z_j) = 1 - \exp\left(-\frac{\Delta^2z}{2\sigma_z^2}\right),
\end{equation}
$\Delta z = z_i-z_j$, i.e. a constant weighting with a Gaussian dip of width $\sigma_z$, which down-weights close pairs in redshift $z_i$. $\sigma_z$ corresponds to the photometic redshift uncertainty, which is typically larger than the intrinsic alignment correlation length. \citet{2002A&A...396..411K} quantify the loss of cosmological information by discarding galaxy pairs and the residual contamination using a quadratic model motivated by conventional tidal torquing. They conclude that the alignment contamination can be effectively suppressed, that the observed correlation function looses amplitude by the weighting, and that the increase in Poisson-noise due to down-weighting can be as large as 25\% for large $\sigma_z$, emphasising the importance of good photometry.

Reduction of intrinsic alignment contamination in the context of the COMBO-17 survey, a medium-deep photometric survey with median redshift $z_\mathrm{med}\simeq0.6$, is the topic of the papers by \citet{2003MNRAS.339..711H} and \citet{2004MNRAS.347..895H}: They pursue the strategy of downweighting galaxy pairs at small separation, and improve on \citet{2002A&A...396..411K} by deriving the optimal pair weight for a spectroscopic survey, which minimises the errors in the weak lensing correlation functino due to intrinsic alignment, assuming a specific alignment model \citep{2000MNRAS.319..649H, 2002MNRAS.335L..89J}, and due to shot noise. They show that the intrinsic alignment contamination can be reduced to the level of pure shot noise for a shallow spectroscopic survey, and that the method applies to photometric surveys like COMBO-17 as well. In a further step, they are able to constrain the alignment model from data, measure the amplitude of the intrinsic ellipticity correlation function and quantify its impact on the measurement of the cosmological parameters $\Omega_m$ and $\sigma_8$ from COMBO-17.

\citet{2003A&A...398...23K} demonstrate that in a tomographic measurement with accurate photometry the intrinsic alignment contamination can be disentangled statistically from the lensing signal, and extend their study to show that the degeneracy between $\Omega_m$ and $\sigma_8$, which is inherent to gravitational lensing and which can be partially lifted using redshift estimates for the source galaxies, can be dealt with even if intrinsic alignments are present. Technically, they prepare artificial data consisting of the weak lensing spectrum, generated from the CDM spectrum assuming a $\Lambda$CDM cosmology, with an intrinsic alignment contamination, using the model taken from \citet{2000MNRAS.319..649H}, and dilute the data by modeling photometry errors in the tomography correlation functions. They continue by carrying out a $\chi^2$-fit to the data with a grid of template functions for the lensing signal as well as for the intrinsic alignments and show that the two contributions can be separated, due to the fact that the intrinsic alignment causes correlations only on short separations and shows a different redshift dependence. In a further study \citet{2005A&A...441...47K} extends the analysis to deal with cross correlations between intrinsic alignment and the weak lensing signal in a linear alignment model, with essentially the same success.

A very elegant solution to the intrinsic alignment problems is proposed by \citet{2008arXiv0804.2292J}: They consider a tomographic setup, where the set of lensed background galaxies are divided into a number of redshift bins and design new line of sight weighting functions which cancel out the shear signal originating from galaxies in the same tomographic bin, by introduction of a line of sight weighting function $B_i(\chi)$, which replaces the redshift distribution of the galaxies $p_i(\chi)$ of conventional lensing tomography. Specifically, the line of sight integral for the weak lensing convergence $\tilde{\kappa}_i$ replaces $\kappa_i$,
\begin{equation}
\tilde{\kappa}_i = \frac{3H_0^2\Omega_m}{2c^2}\int_0^{\chi_H}\dd\chi\:\tilde{G}_i(\chi)\chi\frac{D_+}{a}\delta
\end{equation}
with the changed lensing efficiency $\tilde{G}(\chi)$, 
\begin{equation}
\tilde{G}_i(\chi) = \int_\chi^{\chi_H}\dd\chi^\prime\:B_i(\chi^\prime)\frac{\chi^\prime-\chi}{\chi^\prime},
\end{equation}
which carries out the nulling of the contribution of the large-scale structure inside the tomography bin $i$, by demanding that the comoving distance $\hat{\chi}_i$ the weighting function vanishes, $\tilde{G}(\hat{\chi}_i)=0$, \spirou{assuming locality of the intrinsic alignment effect and should therefore be applicable to the suppression of the $C_{\kappa\epsilon}(\ell)$-contribution as well, despite its longer correlation length}. Quite generally, a loss of information is associated with the nulling, and consequently one would expect a deterioration in the accuracy of cosmological parameters derived from a measurement of the power spectrum $C_{ij}(\ell)$ of the cross-fluctuations in the $\tilde{\kappa}_i$-field. Noticing the analogy to variational problems, \citet{2008arXiv0804.2292J} continue the derivation of the weighting functions $B_i(\chi)$ by constructing the Fisher matrix of the spectra $C_{ij}(\ell)$. The Fisher matrix is defined as the \spirou{expectation value of the second derivative of the logarithmic likelihood with respect pairs of cosmological parameters and provides a measure of the sensitivity of the lensing spectra for constraining the cosmological parameters}. Specifically, the trace of the Fisher matrix, which is equal to $\sum_\mu \sigma_\mu^{-2}$ of the individual errors $\sigma_\mu$ of the cosmological parameters $x_\mu$ due to the Cram{\'e}r-Rao inequality, is maximised, with the nulling condition $\tilde{G}(\hat{\chi}_i)=0$ contained in the spectra. They are able to show both analytically and numerically that the weighting functions $B_i(\chi)$ with the desired properties can be constructed, that the accuracy loss in parameter estimation amounts to typically 30\%, and that the parameter estimates remain unbiased.

\spirou{Considering 3d-lensing measurements, \citet{2008arXiv0801.3270K} quantified the impact of intrinsic alignments along with other uncertainties such as photometry errors on the estimation of cosmological parameters and in particular the dark energy parameters, using an extended Fisher-matrix formalism. They describe the intrinsic alignments with the parameterisation proposed by \citet{2006MNRAS.371..750H} and show that uncorrected intrinsic alignment contamination has a strong impact on the estimation of the dark energy figure of merit $\mathrm{FoM} = [\Delta w_a\Delta w(z_p)]^{-1}$, degrading it by as much as a factor of 2. Additionally, using biased parameters for describing the contamination results in large biases in the estimation of cosmological parameters, but including prior information on the systematics is able to recover the accuracy of the measurement to a large extent.}

Finally, \citet{2007arXiv0710.5171A} provide a general framework for quantifying how systematics in lensing measurements introduce biased values for the estimated cosmological parameters, which can be applied to the intrinsic alignment problem. Their solution to the intrinsic alignment model would be to measure the cosmological parameters from the entire weak lensing data set and correct for the estimation bias due to intrinsic alignment model.

% --- section: summary and outlook --- %
\section{Summary and outlook}\label{sect_summary}
This review article summarises models for galactic angular momenta and angular momentum correlations in the cosmological large-scale structure, and its implications for gravitational lensing. Starting from perturbative models and tidal torquing, angular momentum distributions and correlation functions are derived and simplifications and their applicability discussed. In general, linear tidal torquing is a very successful theory for explaining the angular momentum magnitude and direction, only the nonlinear stages become difficult to describe analytically, as well as the influence of baryonic dynamics. The many functional forms used show the ambiguity in defining an angular momentum correlation function. Interesting future developments might include higher-order statistics of the angular momentum field, which is intrinsically non-Gaussian even if it develops from Gaussian initial fluctuations, or for non-Gaussianities in the initial conditions, either primordial or due to translinear structure formation on small scales. Simulations find correlations between angular momenta and the strong tidal fields of nonlinear structures, illustrating the richness of the cosmological large-scale structure which can not be quantified using 2-point functions alone.

The primary application of angular momentum models are ellipticity correlations, which have been shown to be a serious contaminant of high-precision weak lensing measurements. A plethora of methods has been suggested for removing or suppressing intrinsic ellipticity correlations such that the power of weak lensing surveys for constraining cosmological parameters can be harnessed. Nevertheless one could expect futher development in predicting statistical properties of the intrinsic alignments, especially in comparing linear and quadratic alignment models, and numerical studies of how the ellipticity measured from the luminosity distribution is related to the angular momentum. The suppression and nulling techniques invented to suppress the intrinsic alignment can of course be used to achieve an amplification of the alignment signal over the weak lensing, and it might be possible to combine higher order image distortions such as weak flexions, which should be free of intrinsic alignments, with the weak shear signal, enhancing the understanding of intrinsic and weak lensing induced ellipticity correlations. \spirou{Insight into the alignment of the stellar light distribution with respect to the dark matter host structure could possibly be gained using strong lensing \citep[an overview is provided by][]{2006glsw.conf.....M}, by measuring the halo's potential well and combining with kinematical data.}

% --- section: Acknowledgements --- %
\section*{Acknowledgements}
I would like to thank Rob Crittenden and Matthias Bartelmann for valuable comments on angular momentum correlations. I am grateful to Catherine Heymans and Martin Kilbinger for sharing their expertise on gravitational lensing and for suggestions on the draft. I am indepted to Nabila Aghanim for giving me the opportunity to work on this review, and to Sarah Bridle for organising an excellent workshop on intrinsic alignments. I would like to thank in particular C. Porciani, O. Hahn, R. Crittenden and C. Heymans, who provided figures from their works for illustration and Blackwell Publishing as well as the American Astronomical Society for granting me permission to reproduce the figures, which appear in the journals {\em Monthly Notices of the Royal Astronomical Society} and the {\em Astrophysical Journal}, respectively. I am grateful for the comments of the anonymous referee, which led to a sigificant improvement of the review. My work is supported by an STFC postdoctoral fellowship.

% --- section: SZ definitions --- %
\bibliography{bibtex/aamnem,bibtex/references}
\bibliographystyle{mn2e}

\appendix

\bsp

\label{lastpage}

\end{document}